\documentclass[aps,pra,showpacs,notitlepage,floatfix,twocolumn,superscriptaddress]{revtex4-1}
\usepackage{bbm}
\usepackage{mathrsfs}
\usepackage{epsfig}
\usepackage{graphicx}
\usepackage{amsfonts}
\usepackage[figuresright]{rotating}
\usepackage{amssymb}
\usepackage{amsmath}
\usepackage{dcolumn}
\usepackage{bm}
\usepackage{braket}
\usepackage[colorlinks,linkcolor=blue,anchorcolor=blue,citecolor=blue,urlcolor=blue]{hyperref}

\def\be{\begin{equation}} \def\ee{\end{equation}}
\def\bea{\begin{equation}\begin{aligned}} \def\eea{\end{aligned}\end{equation}}

\def\dxy{d_{xy}/d_{x^2-y^2}}

\graphicspath{{./}{./figure/}}

\begin{document}
\title{Orbital-Active Dirac Materials from the Symmetry Principle}

\author{Shenglong Xu}
\affiliation{Department of Physics \& Astronomy, Texas A\&M University, College Station, Texas 77843, USA}
\author{Congjun Wu}
\email{wucongjun@westlake.edu.cn}
\affiliation{Department of Physics, School of Science, Westlake University, Hangzhou 310024, Zhejiang, China}
\affiliation{Institute for Theoretical Sciences, Westlake University, Hangzhou 310024, Zhejiang, China}
\affiliation{Key Laboratory for Quantum Materials of Zhejiang Province, School of Science, Westlake University, Hangzhou 310024, Zhejiang, China}
\affiliation{Institute of Natural Sciences, Westlake Institute for Advanced Study, Hangzhou 310024, Zhejiang, China}

\begin{abstract}
Dirac materials, starting with graphene, have drawn tremendous research interest in the past decade. 
Instead of focusing on the $p_z$ orbital as in graphene, we move a step further and study orbital-active Dirac materials, where the 
orbital degrees of freedom transform as a two-dimensional irreducible representation of the lattice point group. 
Examples of orbital-active Dirac materials occur in a broad class 
of systems, including transition-metal-oxide heterostructures, 
transition-metal dichalcogenide monolayers, germanene, 
stanene, and optical lattices. 
Different systems are unified based on symmetry principles. 
The band structure of orbital-active Dirac materials features Dirac 
cones at $K(K')$ and quadratic band touching points at $\Gamma$, 
regardless of the origin of the orbital degrees of freedom. 
In the strong anisotropy limit, i.e., when the $\pi$-bonding can be neglected, flat bands appear due to the destructive interference. 
These features make orbital-active Dirac materials an even wider 
playground for searching for exotic states of matter, 
such as the Dirac semi-metal, ferromagnetism, Wigner crystallization, quantum spin Hall state, and quantum anomalous Hall state.
\end{abstract}
\maketitle


\section{Introduction}
Graphene opened up a new era of topological materials, followed by the discovery of topological insulators, topological superconductors, and semi-metals in both two and three dimensions (See reviews~\cite{Hasan2010b, Qi2011, Armitage2018}). 
Since then, the interplay between topology and correlation has been the primary focus of condensed matter research. 
Graphene and its variants, due to its excellent electronic and 
mechanical properties~\cite{Neto2007, DasSarma2011}, have become 
wonderful platforms for hosting exotic phases of matter and also find themselves widely applicable in electric device engineering and 
material science.
The characteristic feature of graphene is the appearance of Dirac cones 
in the spectrum, tied to the symmetry of the underlying honeycomb lattice. 
Two sublattices ($A$ and $B$) of the honeycomb lattice transform into each 
other under the simplest non-abelian point group $C_{3v}$, 
which contains 3-fold rotations and in-plane reflections. 
At the $K (K')$ point of the Brillouin zone, the wavefunctions of 
$A$ and $B$ sublattices form the two-dimensional ($E$) irreducible representations (irrep) of the $C_{3v}$, enforcing the Dirac cones. 
Once there, the Dirac cones are stable as long as 
time-reversal and inversion symmetries are preserved.

The on-site $p_z$ orbital of graphene transforms trivially 
(it belongs to the $A_1$ irrep)  under the site symmetry group 
$C_{3v}$.  
It is natural to ask what happens if the on-site orbitals 
form the $E$-irrep of the point group.  
The $E$-irrep features the double degeneracy and anisotropy,
which is expected to bring rich orbital physics in graphene-like 
Dirac materials. 
Such a situation arises in many distinct systems.
It was initially studied in optical lattices, where the two-dimension 
irrep is realized by the $p_x$ and $p_y$ orbitals in the harmonic trap~\cite{Wu2007, Wu2008}. 
In transition-metal-oxide heterostructures~\cite{Xiao2011, Ruegg2011, Ruegg2012,Yang2011a} and transition-metal-dichalcogenide monolayers~\cite{Qian2014}, the $d$-orbitals decompose based on 
the $C_{3v}$-symmetry and are active near the Fermi surface.  
In the hexagonal monolayers of heavy elements, such as  Germanene, 
Stanene, and Bismuthene, the $(p_x, p_y)$ doublets realize the 
orbital degrees of freedom.
Due to the enriched orbital structure of the Dirac cone, the gap
opening, which turns out to be topologically non-trivial, 
equals to the atomic spin-orbit coupling, hence, it can be
very large reaching the order of 1eV 
~\cite{Xu2013, Wu2014, Zhang2014, Reis2017,xia2021high,jin2022large}.

Even in simple carbon systems, orbital physics can be realized via lattice engineering, for example, organic framework~\cite{Wang2013, Wang2013a} and graphene-kagome lattice~\cite{Chen_kagome2018}.   
Remarkably, recently experiments~\cite{Cao2018, cao2018correlated, yankowitz2019tuning} on twisted bilayer graphene revealed  Mott 
insulator and superconductivity phases, and it is proposed that 
the low-lying degrees of freedom are compatible with two 
orbitals on the honeycomb lattice as well~\cite{Po2018, Yuan2018, Liu2018,venderbos2018correlations,dodaro2018phases,fidrysiak2018unconventional}.
Furthermore, the orbital degrees of freedom do not have to be 
electronic and can manifest themselves as the polarization modes 
of polaritons in photonic lattices~\cite{Jacqmin2014, Milicevic2017} 
and phonons in graphene and mechanical 
structures~\cite{Zhang2015, Roman2015, Stenull2016, Zhu2018}.

Given all these interconnected systems and the increasing realizations of orbital-active Dirac materials, this work aims to bridge all the different systems through the symmetry principle. 
Despite the vastly different origins, the orbital degrees of freedom 
can be understood as the irreducible representations of the site 
symmetry of the lattice, which leads to universal properties. 
We show that the symmetry alone enforces the Dirac cone at $K (K')$ 
point and the quadratic band touching at the $\Gamma$ point.
Various gap opening mechanisms and interaction effects are discussed, 
which lead to the quantum spin Hall effect and quantum anomalous 
Hall effect. 
In particular, when the $E_g$ doublets realize orbital degrees of 
freedom, the resulting topological insulator states carry octupole order.
Finally, the method employed here for studying the doubly-degenerate orbitals in the honeycomb lattice can be readily generalized to orbital degrees of freedom arising from larger lattice point groups.

The rest of the paper is organized as follows. 
In  Sec.~\ref{sec:honeycomb_orb}, the symmetry of 
the honeycomb lattice and the orbital realization of the 
on-site irreducible representations are studied by
focusing on the $d$-orbitals.
In Sec.~\ref{sec:band_structure}, the band structure of the orbital-active honeycomb lattice systems is derived from a simple tight-binding model. 
In Sec.~\ref{sec:general_symmetry}, we go beyond the simple tight-binding model and  demonstrate that many interesting features of the band structure are solely protected by the lattice symmetry. 
In Sec.~\ref{sec:gap_opening}, various band gap opening mechanisms are stuided. 
In Sec.~\ref{sec:interaction}, the interplay between band structure 
and the interaction effects is discussed. 
Sec.~\ref{sec:summary} is left for summary and outlook.

\section{The honeycomb lattice and orbital symmetries}
\label{sec:honeycomb_orb}

\begin{figure}
\includegraphics[height=0.45\columnwidth, width=0.45\columnwidth]
{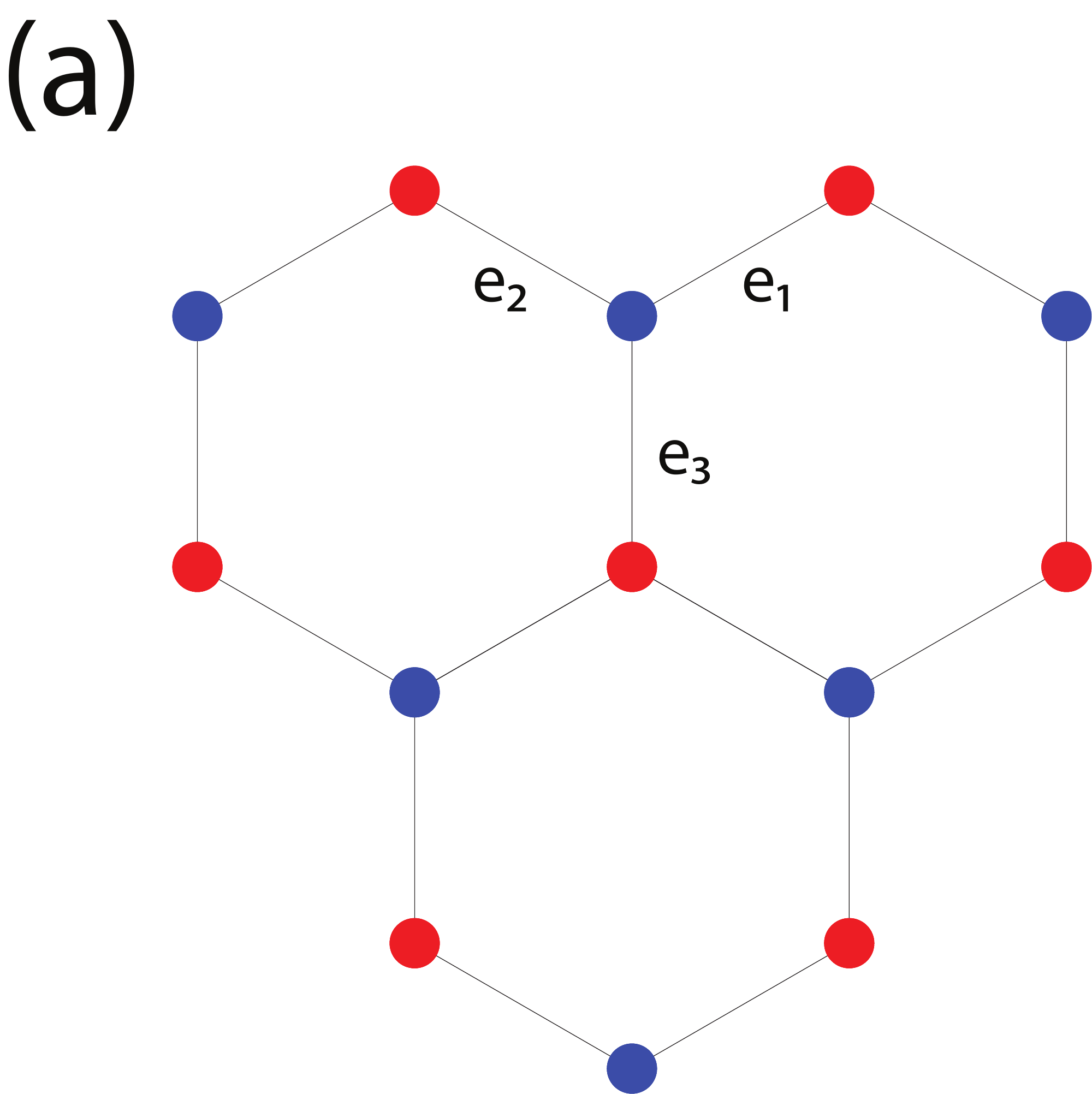}
\includegraphics[height=0.45\columnwidth, width=0.45\columnwidth]
{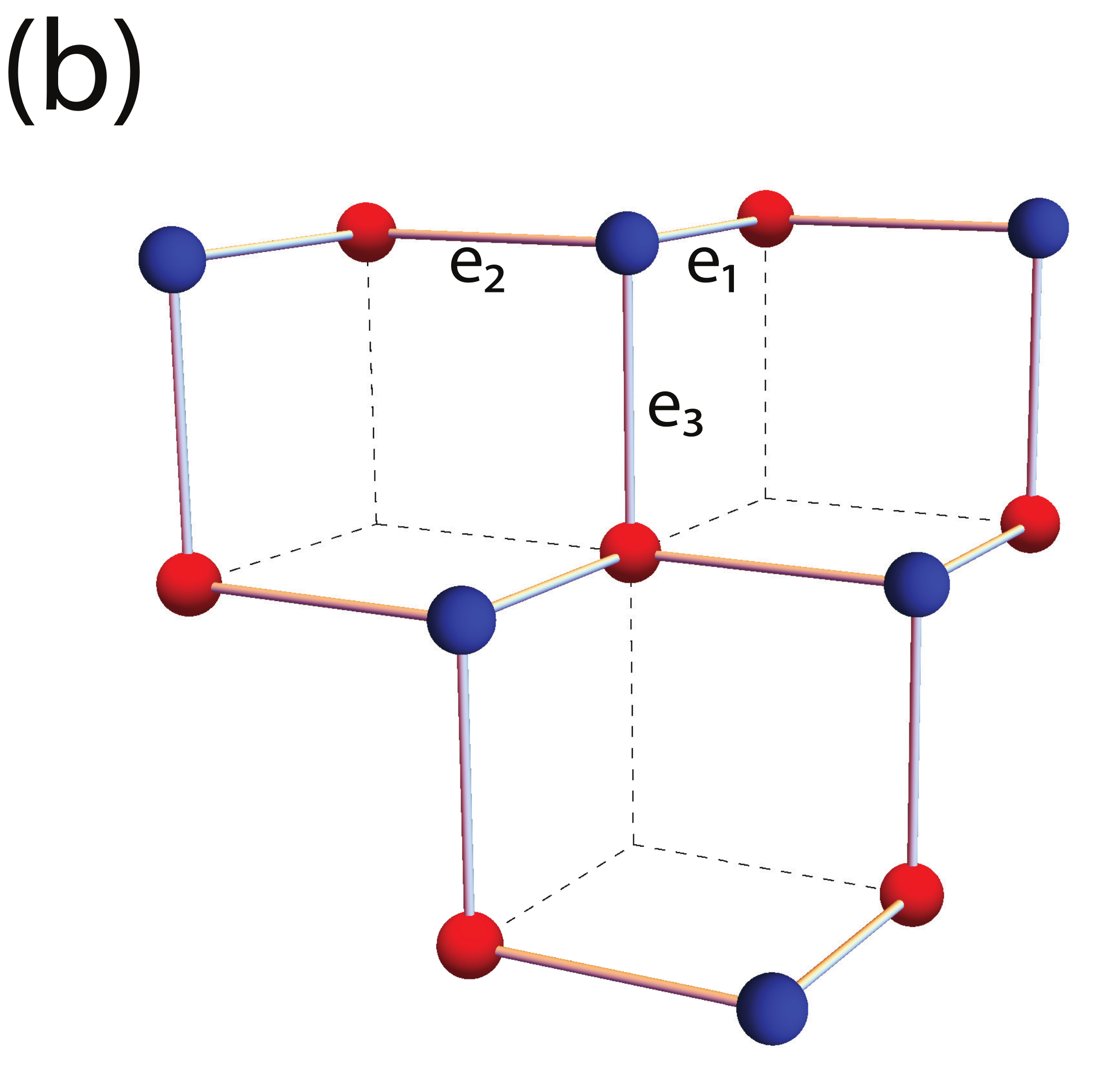}
\caption{(a) The standard honeycomb lattice with $C_{6v}$ point symmetry group.   (b) The buckled  honeycomb lattice with the symmetry group is downgraded to $D_{3d}$.}
\label{fig:lattice}
\end{figure}

We start with reviewing the symmetry of the planar honeycomb lattice.
The planar honeycomb lattice, sketched in Fig.~\ref{fig:lattice}(a),  
consists of two sublattices $A$ (blue) and $B$ (red). The three nearest neighbor vectors are labeled as $\hat{e}_1\sim \hat{e}_3$.
The symmetry of the lattice is described by the space group $P6mm$, a direct product of the
point group $C_{6v}$ and the translation symmetry of the triangular
Bravis lattice \footnote{If one also considers the mirror symmetry taking $z$ to $-z$, the point group is $D_{6h}$ and the space group is $P6/mmm$.}. The maximal point group $C_{6v}$ is realized at
the centers of the hexagons.
On the other hand,  the point group symmetry acting on a lattice site, called site symmetry, is a subgroup of the maximal point group. The site symmetry group is important because it affects the  orbital part of the wavefunction of the degrees of freedom living on lattice sites (such as electrons, phonons, etc.)
The site symmetry of the honeycomb lattice is $C_{3v}$ generated by a 3-fold rotation axis and
three vertical reflection planes (e.g., the $yz$-plane and its
symmetry counterparts by rotations of $\pm 120^\circ$).
In contrast, the reflection with respect to the $xz$-plane and its
symmetry counterparts by rotations of $\pm 120^\circ$ interchange
the $A$ and $B$ sublattices and are not included in the site symmetry.

The orbital of the onsite degrees of freedom is classified by the irreducible representations of the site symmetry group.
The $C_{3v}$ group has three irreducible representations (irrep), including two 1d irreps $A_{1,2}$ and a 2d irrep $E$ as explained in Appendix~\ref{appendix:group}. 
The irreps fully determine the symmetry structure of the onsite 
degrees of freedom, regardless of their microscopic origins.
In this article, we focus on electron atomic orbitals. 
Taking $z$ axis perpendicular to the lattice plane, the $s$ and $p_z$ 
orbitals realize the $A_1$-irrep and lead to the remarkable electronic structure of graphene. 
In contrast, the $p_x$ and $p_y$-orbitals realize the two 
dimensional $E$-irrep. 
This doublet can also be organized into the complex basis $p_x \pm ip_y$
which are eigenstates of the orbital angular momentum $L_z$ 
with eigenvalues $\pm 1$, respectively.
As to the 5-fold $d$-orbitals, the $d_{r^2-3z^2}$ falls into the $A_1$ irrep.
The remaining four form two $E$ irreps: the $(d_{xz}, d_{yz})$ 
doublet and the $(d_{xy}, d_{x^2-y^2})$ doublet.
The complex orbitals $d_{xz}\pm i d_{yz}$, and 
$d_{xy}\pm d_{x^2-y^2}$ carry orbital angular momentum 
numbers $\pm 1$ and $\mp 2$, respectively.
Since the site symmetry group only has one 2d irrep, the three doublets, $(p_x, p_y)$, $(d_{xz}, d_{yz})$ and $(d_{xy}, d_{x^2-y^2})$ 
are equivalent as far as the symmetry is concerned. 
One can explicitly check that the group elements of 
the site symmetry $C_{3v}$ have the same matrix representation
of the $E$-irrep.

A closely related lattice structure sketched in Fig~\ref{fig:lattice}($b$) 
is dubbed buckled honeycomb lattice, which can be viewed as a bilayer 
of sites taken from a cubic lattice in the $(1,1,1)$ direction.
The blue and red dots form a honeycomb lattice when projecting 
into the $(1,1,1)$ plane.
Compared to the planner honeycomb lattice, the point group symmetry of the buckled lattice downgrades from $d_{6h}$ to $d_{3d}$, where the six-fold rotation becomes a rotoreflection. 
On the other hand, the site symmetry remains the same, described by $C_{3v}$. As a result, based on previous analysis, the realizations of the $E$-irrep 
in the buckled lattice must be equivalent to the $(p_x, p_y)$ doublet in the planar case. 
Here we focus on the $d$-orbitals and establish this equivalence. 
The buckled honeycomb lattice originates from the cubic lattice. 
Taking the $z$-axis along the $(0,0,1)$-direction, the 5-fold 
$d$-orbitals split into a $T_{2g}$ triplet $(d_{yz}, d_{zx}, d_{xy})$ 
and an $E_g$ doublet $(d_{x^2-y^2}, d_{r^2-3z^2})$, which are irreps 
of the $O_h$ point group.
The site symmetry of the buckled lattice $C_{3v}$ is a subgroup 
of $O_h$. 
The $E_g$ doublet falls into the only 2d irrep $E$ of $C_{3v}$, 
while the $T_{2g}$ triplet further splits into the 1d irrep $A_1$ 
and the 2d irrep $E$.

To make the connection between the $E_g$ doublet and the orbital 
realization of the $E$ irrep in the planar case more explicit, 
we rotate the frame of the buckled lattice so that the $z$-axis 
is along the 3-fold axis $(1,1,1)$. 
Then the $E_g$ doublet becomes
\bea
d_{x^2-y^2}&\rightarrow \frac{1}{\sqrt{3}}(d_{xy}+\sqrt{2}d_{xz}), \\
d_{r^2-3z^2}&\rightarrow  \frac{1}{\sqrt{3} }(d_{x^2-y^2}+\sqrt{2} d_{yz}).
\label{eq:rotated_eg}
\eea
Hence, the $E_g$ orbitals are a superposition of two
$E$ doublets $(d_{xy}, d_{x^2-y^2})$ and $(d_{xz}, d_{yz})$ in the planar case.
Therefore, as far as the site symmetry $C_{3v}$ is concerned, the $e_g$ doublet is equivalent to the
$(p_x, p_y)$ doublet in the planar case.
In fact, the mapping can be made explicit as
\bea
d_{x^2-y^2} \leftrightarrow p_x, \ \ \,
d_{r^2-3z^2} \leftrightarrow p_y.
\eea

For completeness,  the decomposition of the five $d$-orbitals into two
$E$ irreps and one $A_1$ irrep of the $C_{3v}$ group is presented 
as follows,
\bea
\begin{cases}
(d_{x^2-y^2}, d_{r^2-3z^2})  & E  \\
\left (\frac{1}{\sqrt{2}}(d_{yz} -d_{zx}), \frac{1}{\sqrt{6}} (d_{yz}+d_{zx}-2d_{xy})\right )  & E \\
\frac{1}{\sqrt{3}}(d_{xy}+d_{yz}+d_{zx}) &A_1, \\
\end{cases}
\label{eq:111_irrep}
\eea
choosing $(1,1,1)$ as the rotation axis.
In addition to the $E_g$ orbitals which become an $E$-representation,
the $T_{2g}$-orbitals split into one $E$ irrep and one
$A_1$ irrep.
In principle, the two $E$-representations froming the $E_g$ and $T_{2g}$
orbitals can mix.
In transition-metal-oxides where the buckled lattice is relevant, there 
is often an oxygen octahedron around each transition metal ion.
The octahedron introduces a large crystal field that splits the $E_g$ 
and $T_{2g}$ orbitals.
Hence, the mixing between the $E$ irrep of $C_{3v}$ derived 
from the $E_g$ orbitals and that from the $T_{2g}$ orbitals is weak.

\section{Magnetic octupole moment of the $E_g$ doublet}
\label{sec:octupole}
Although all realizations of $E$-irrep of $C_{3v}$ are equivalent 
from the symmetry consideration. 
The $E_g$ orbitals are special physically and worth special attention. 
The key difference lies in the angular momentum of the complex 
combination of the doublets. 
In the case of $(p_x, p_y)$, the complex combination $p_x \pm i p_y$ 
takes the form $\exp(\pm i \theta)$, and thus carries angular 
momentum $\pm 1$ along the rotating axis.  
The same applies to the $(d_{xz}, d_{yz})$ doublet. 
In the case of $(d_{xy}, d_{x^2-y^2})$, the complex combination 
$d_{xy} \pm i d_{x^2-y^2}$ takes the form $\exp (\mp i 2\theta )$ 
and thus carries angular momentum $\mp 2$.
In contrast, the angular momentum of the complex combination of the $E_g$ orbitals $d_{x^2-y^2}\pm d_{r^2-3z^2}$ vanishes.  
From Eq.~\eqref{eq:rotated_eg},  the complex combination of $E_g$ doublet 
can be viewed as the weighted superposition of the complex combinations 
of the $(d_{xz}, d_{yz})$ and $(d_{xy}, d_{x^2-y^2})$.
The angular momentum of the two doublets cancel each other, leading to the zero angular momentum of the $E_g$ doublet.

Instead of the angular momentum, the complex $E_g$ orbitals carry 
higher rank magnetic moment,  measured by spherical tensor 
operators $Y_{lm}$. 
A list of spherical tensor operators constructed from the angular 
momentum operator $\vec L$ in the $d$ orbital space can be 
found in appendix~\ref{appendix:tensor}.
Going through all the higher rank tensor operators, we find that 
the leading non-vanishing spherical tensor operators projecting 
into $E_g$ orbital is
\bea
P_{E_g} Y_{3,\pm 2}P_{E_g}=\mp 3\sqrt{\frac{5}{2}}  i\sigma_2,
\eea
where $P_{E_g}$ is the projection operator. 
Two non-vanishing components of the rank-3 spherical tensor operators 
can be grouped into a single cubic harmonic tensor 
$\hat f_{xyz}=\frac{i}{\sqrt 2}(Y_{2,-2}-Y_{2,2})$.
It is projected into $E_g$ orbital space as
\bea
P_{E_g}\hat f_{xyz} P_{E_g}=-3\sqrt 5 \sigma_2.
\eea
where $\hat f_{xyz}$ corresponds to the octupole magnetic moment. 
Therefore, the complex combinations of the $E_g$ orbital, instead of 
carrying angular momentum, carry octupole magnetic moment, 
which was proposed to be the ``hidden order"  
in certain strongly-correlated electronic systems \cite{Santini2000,Brink2001,Kuramoto2009,Jackeli2009,Li2016}.

\section{The band structure of the orbital active honeycomb lattice}
\label{sec:band_structure}
In this section, we study the band structure of the orbital active
honeycomb lattice, including the planer and the buckled ones. 
To be concrete, we first introduce a simple nearest neighboring 
tight-binding model before presenting more general scenarios 
in the next section.

\subsection{Constructing the tight-binding model}
\begin{figure}
\includegraphics[height=0.32\columnwidth, width=0.32\columnwidth]
{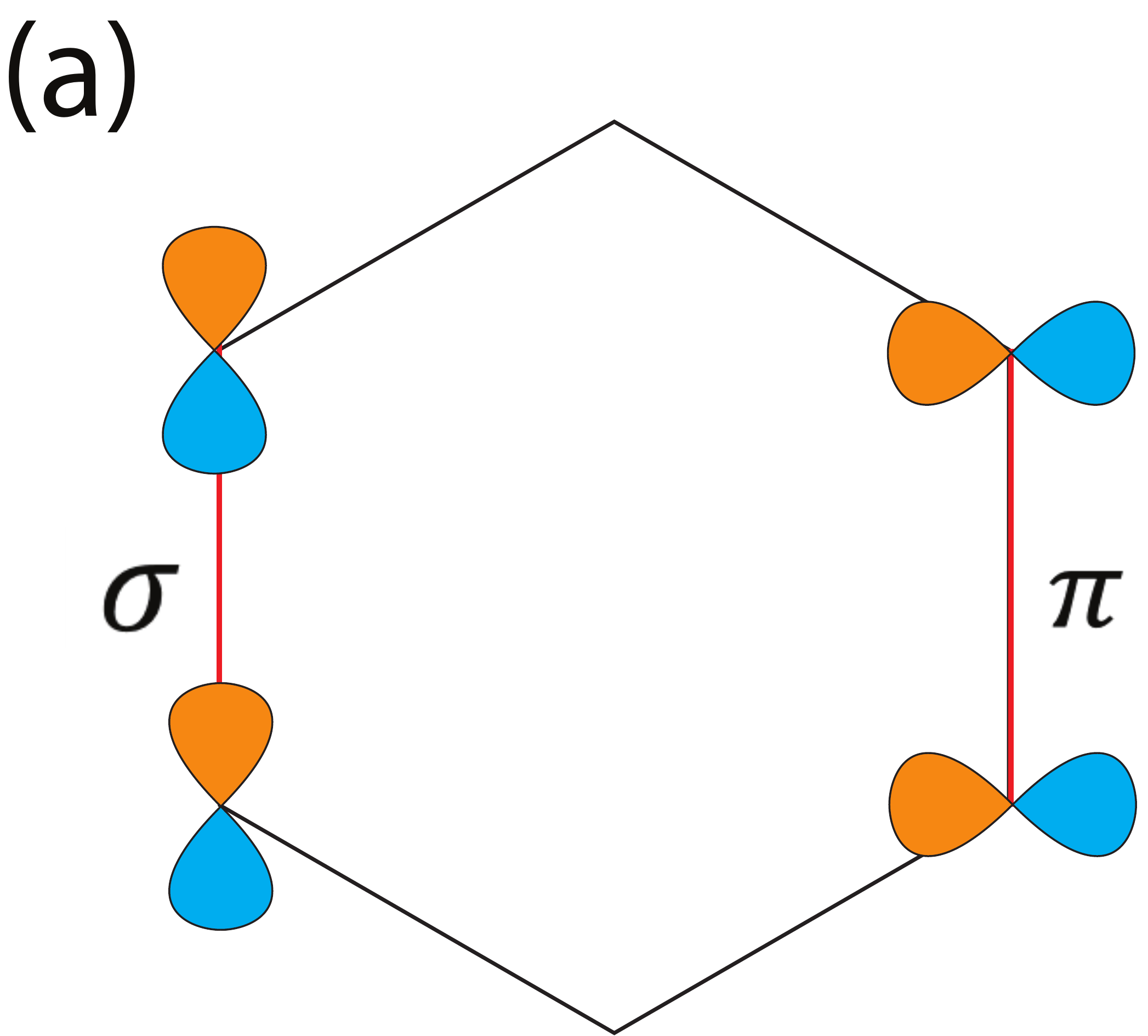}
\includegraphics[height=0.32\columnwidth, width=0.32\columnwidth]
{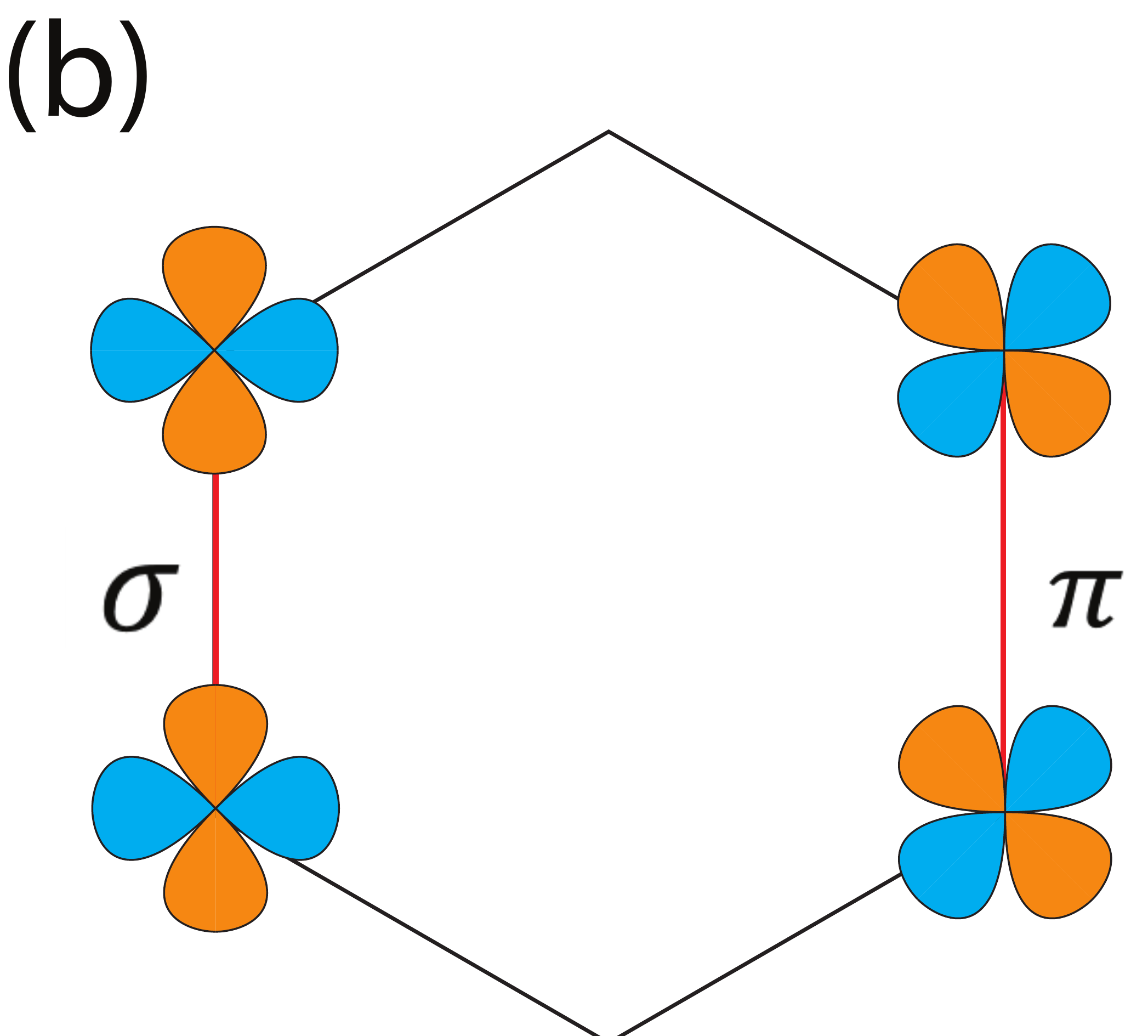}
\includegraphics[height=0.32\columnwidth, width=0.32            \columnwidth]
{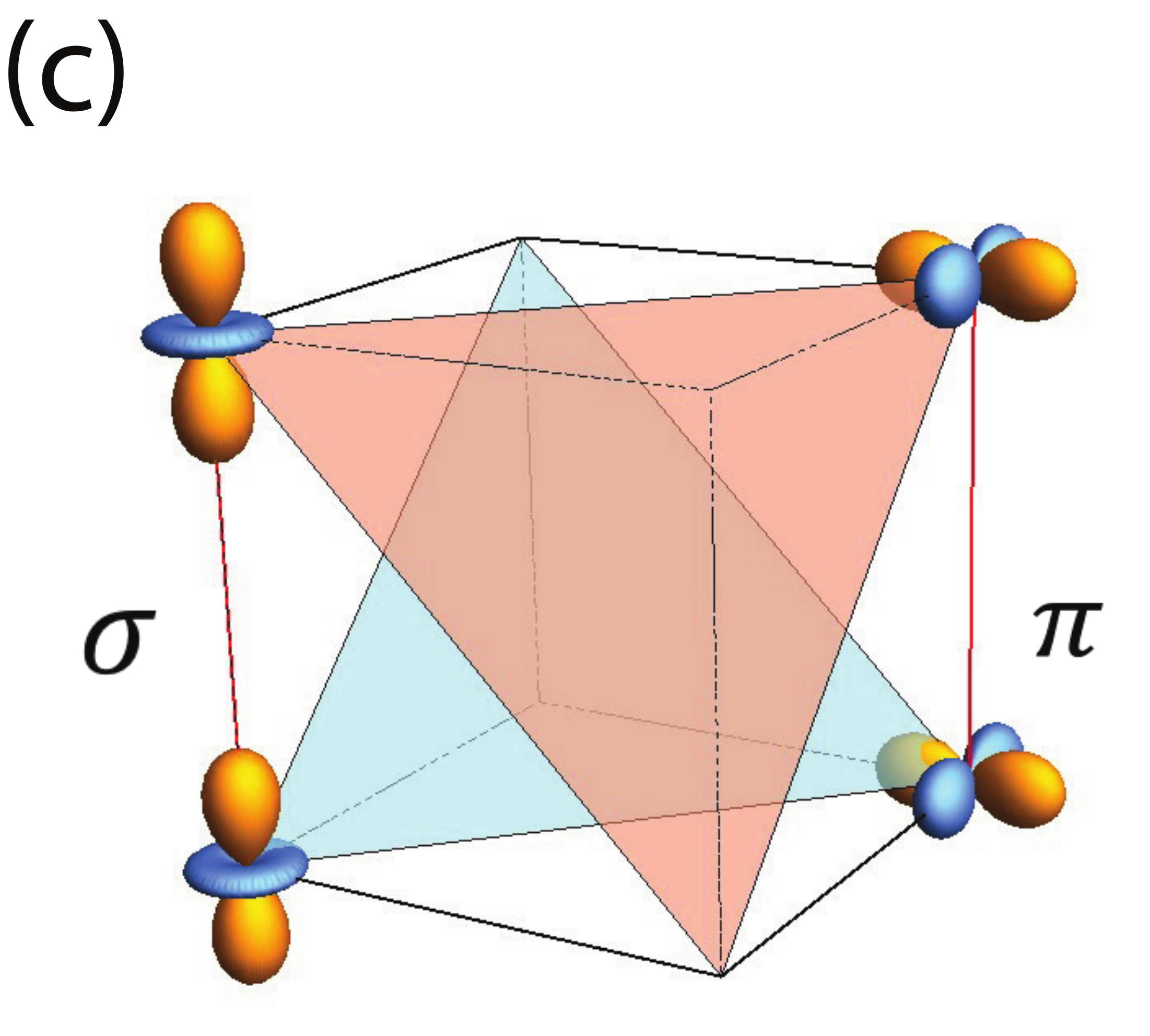}
\caption{The orbital configuration of the $\sigma$-bonding on the $\hat{e}_3$ bond for (a) $(p_x, p_y)$ doublet; (b) $(d_{xy}, d_{x^2-y^2})$ doublet; (c) $E_g$ doublets.}
\label{fig:bonding}
\end{figure}

In the orbital active honeycomb lattice, the hopping between 
neighboring sites occurs between different components of 
the $E$-irrep and thus is more complicated than the orbital 
inactive case.
There are two kinds of hopping processes allowed by symmetry 
on a bond.
In terms of the chemistry convention, the amplitude of the
$\sigma$-bonding is much larger than that of the $\pi$-bonding.
The difference in the hopping amplitude arises from the 
anisotropy of the orbital wavefunction.
The bonding direction of the $\sigma$ ($\pi$)-bond is along the 
direction of the maximal (minimal) angular distribution of 
orbital wavefunctions.
The $\sigma$-bond and the $\pi$-bond for all orbital realizations of 
the $E$-irrep are shown in Fig.~\ref{fig:bonding} for one of nearest neighboring bond $\hat{e}_3$. 
In the planar case, the $\sigma$-bonding orbital is $p_y$, 
$d_{yz}$ or $d_{x^2-y^2}$, and in the buckled case, 
the $\sigma$ bonding orbital is $d_{r^2-3z^2}$.
The $\sigma$ bonding orbitals along other nearest neighboring 
bonds are linear combinations of the two orbitals in the $E$ irrep, 
obtained from applying 3-fold rotation on $p_y$, $d_{yz}$, 
$d_{x^2-y^2}$ or $d_{r^2-3z^2}$. 

Since all the different doublets form the same irrep of 
the $C_{3v}$ group, they transform in the same way under rotation.
To unify the notation, we use $\gamma_{x,y}$ to represent 
the two states in the $E$ irrep for different orbital realizations,
where $\gamma_x$ stands for $p_x$, $d_{xy}$ or $d_{x^2-y^2}$, and 
$\gamma_y$ stands for $p_y$, $d_{x^2-y^2}$ or $d_{r^2-3z^2}$, 
correspondingly.
$\gamma_1,\gamma_2,\gamma_3$ are defined to be the $\sigma$-bonding 
orbitals along the three nearest neighboring bonds 
$\hat{e}_1, \hat{e}_2, \hat{e}_3$, respectively,
\bea
\gamma_1 = \frac{\sqrt{3}}{2} \gamma_x +\frac{1}{2} \gamma_y, 
\ \ \gamma_2 =- \frac{\sqrt{3}}{2} \gamma_x +\frac{1}{2}\gamma_y,  
\ \ \gamma_3 = - \gamma_y.
\eea

Since the $\sigma$ bonding is much stronger than the $\pi$ bonding, we neglect the $\pi$ bonding and construct the single particle Hamiltonian of the nearest neighboring $\sigma$ bonding. Using  $\gamma_1\sim\gamma_3$, the Hamiltonian can be conveniently written as
\bea
H_0=t_\parallel \sum_{\vec{r}\in A,j=1,2,3}
\Big\{\gamma_j^\dagger (\vec{r}+\hat e_j) \gamma_j(\vec{r}) +h.c. \Big\}.
\label{eq:tight-binding}
\eea
where the summation over $\vec{r}$ is only on the A sublattice and $\hat{e}_1\sim \hat{e}_3$ are the unit vectors pointing from A site to its three nearest neighboring B sites on the planar honeycomb lattice
\bea
\hat{e}_1&=&\frac{\sqrt{3}}{2} \hat{e}_x +\frac{1}{2} \hat{e}_y, 
\ \ \
\hat{e}_2 =-\frac{\sqrt{3}}{2} \hat{e}_x +\frac{1}{2}\hat{e}_y, 
\ \ \
\hat{e}_3=-\hat{e}_y.
\eea
The nearest neighboring distance is set to 1.
In the case of the buckled honeycomb lattice, the three nearest 
neighboring vectors are the same as in the planar case 
when the coordinates are projected onto the $(1,1,1)$ plane.

The Hamiltonian has the same form for different realizations of the 
$E$-irrep of the site symmetry $C_{3v}$ for both the planar 
and buckled honeycomb lattice.
This demonstrates the power and elegance of the symmetry principle.

\subsection{The spectra and wavefunctions}

\begin{figure}[h]
\includegraphics[width=0.45\columnwidth]
{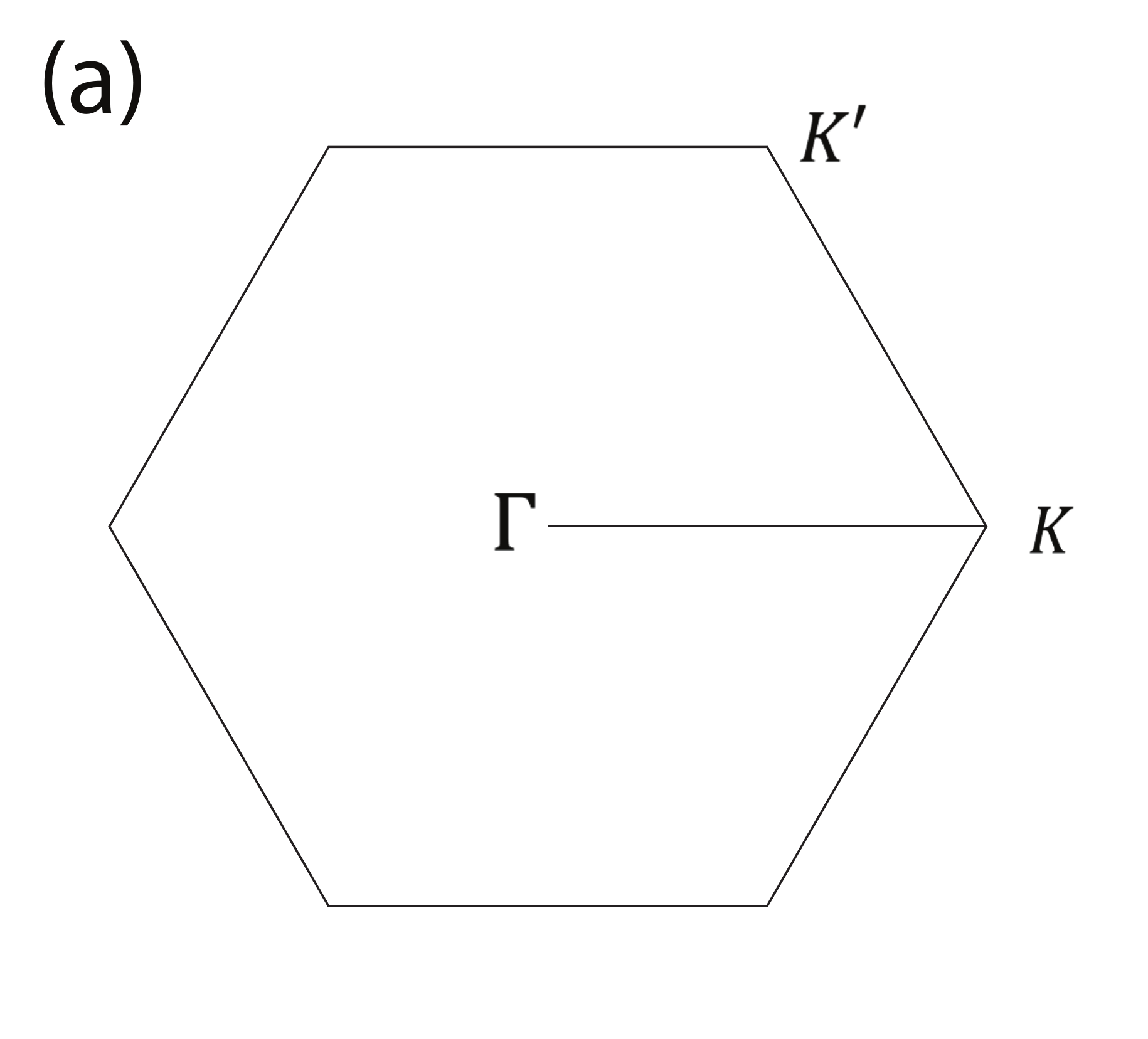}
\includegraphics[width=0.45\columnwidth]
{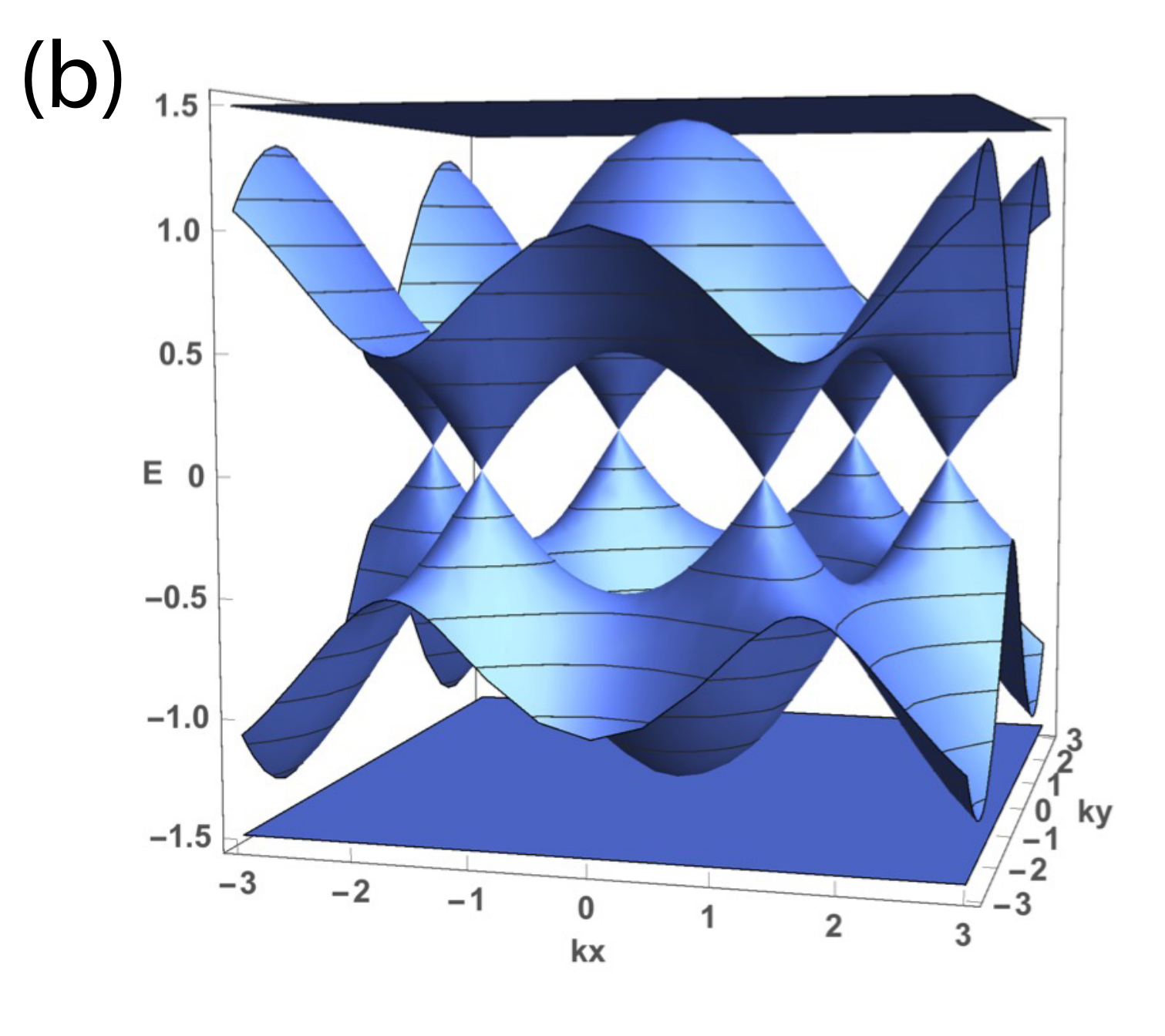}
\caption{(a) The Brillouin zone of the honeycomb lattice. (b) The band structure of the tight binding model is described in Eq.~\eqref{eq:tight-binding} in the strong anisotropy limit. There are four bands (The spin degrees of freedom add another copy and are omitted). The bottom band and the top band are completely flat, while the middle two bands have the same dispersion relations as in graphene.}
\label{fig:spectrum}
\end{figure}

The Hamiltonian Eq.~\eqref{eq:tight-binding} is ready to be 
diagonalized in momentum space, in which a 4-component 
spinor $\psi(\vec k)$ is defined as
\bea
\psi(\vec k)= (\gamma_{x, A}(\vec k),\gamma_{y,A}(\vec k), \gamma_{x,B}(\vec k), \gamma_{y,B}(\vec k))^T,
\eea
where $A$ and $B$ refer to the two sublattices. The annihilation  operators $\gamma_{x,y}(k)$  is defined as
\bea
\gamma_{x, y}(\vec k)&= \frac{1}{\sqrt{N}}\sum\limits_{\vec{r}} \gamma_{x,y}(\vec{r})e^{-i \vec  k\cdot \vec{r}}.
\eea
The crystal momentum $\vec k$ is defined in the Brillouin 
zone shown in Fig.~\ref{fig:spectrum}(a). 
In the case of the buckled honeycomb lattice, 
$\vec{r}$ is the projected coordinate in the $(1,1,1)$ plane.

With this setup, the Hamiltonian takes the following block form:
\bea
H(k)=\begin{pmatrix}
0   &H_{AB}\\
H_{AB}^\dagger  & 0 \\
\end{pmatrix},
\label{eq:H_k}
\eea
where
\bea
H_{AB}= t_\parallel
\begin{pmatrix}
\frac{3}{4} (e^{i \vec k\cdot \hat e_1}+e^{i\vec k\cdot \hat e_2})
& \frac{\sqrt 3}{4} (e^{i \vec k\cdot \hat e_1}+e^{i \vec k\cdot \hat e_2})  \\[0.9 em]
\frac{\sqrt 3}{4} (e^{i \vec k\cdot \hat e_1}+e^{i\vec k\cdot \hat e_2})
& \frac{1}{4}(e^{i \vec k\cdot \hat e_1}+e^{i \vec k\cdot \hat e_2}+2 e^{i \vec k \cdot
\hat e_3})\\[0.9em]
\end{pmatrix}.
\eea
There are four band. 
The middle two bands exhibit exactly the same dispersion 
as that in graphene:
\bea
E_{2,3}=\mp \frac{t_\parallel}{2} 
\left |\sum\limits_ie^{i \vec k \cdot \hat e_i} \right |
=\mp \frac{t_\parallel}{2} 
\sqrt{3+2 \sum \limits_{i=1}^3\cos \vec  k \cdot\vec b_i},
\eea
where $ \vec b_i =\frac{1}{2} \epsilon_{ijk}(\hat e_j-\hat e_k)$ 
are the next nearest neighboring vectors. 
The bands display two Dirac cones at $K$ and $K'$. 
In addition, Fermi surface nesting and Van Hove singularity 
occur at $1/4$-filling above and below the Dirac point.
The wavefunctions associated with the middle two bands are
\bea
\ket{\psi(\vec k)}_{2,3}=\frac{1}{\sqrt {N_0}}\left (e^{-i\frac{\theta}{2}}\sum\limits_i \hat e_i e^{i \vec k \cdot \hat e_i },\pm e^{i\frac{\theta}{2}}\sum\limits_i \hat e_i e^{-i\vec k \cdot \hat e_i} \right ),
\label{eq:psi_23}
\eea
where the phase $\theta=\text{arg}(\sum\limits_ie^{i \vec k\cdot \hat e_i })$ and the normalization $N_0=6-2  \sum \limits_{i=1}^3\cos \vec k \cdot \vec b_i$.

On the other hand, the top and the bottom bands are perfectly 
flat with the energy,
\bea
E_{1,4}=\mp\frac{3}{2}t_\parallel.
\eea
They connect to the middle two bands at the $\Gamma$ point.
The corresponding wavefunctions are represented as
\bea
\ket{\psi(\vec k)}_{1,4}=\frac{1}{\sqrt{3N_0}}\left (\sum\limits_i  \vec b_i e^{-i \vec k \cdot \hat e_i },\pm \sum\limits_i  \vec b_i e^{i \vec k \cdot \hat e_i} \right ),
\label{eq:psi_14}
\eea
and the energy dispersions are plotted in Fig. \ref{fig:spectrum}.

\begin{figure}
\includegraphics[height=0.35\columnwidth, width=0.9\columnwidth]
{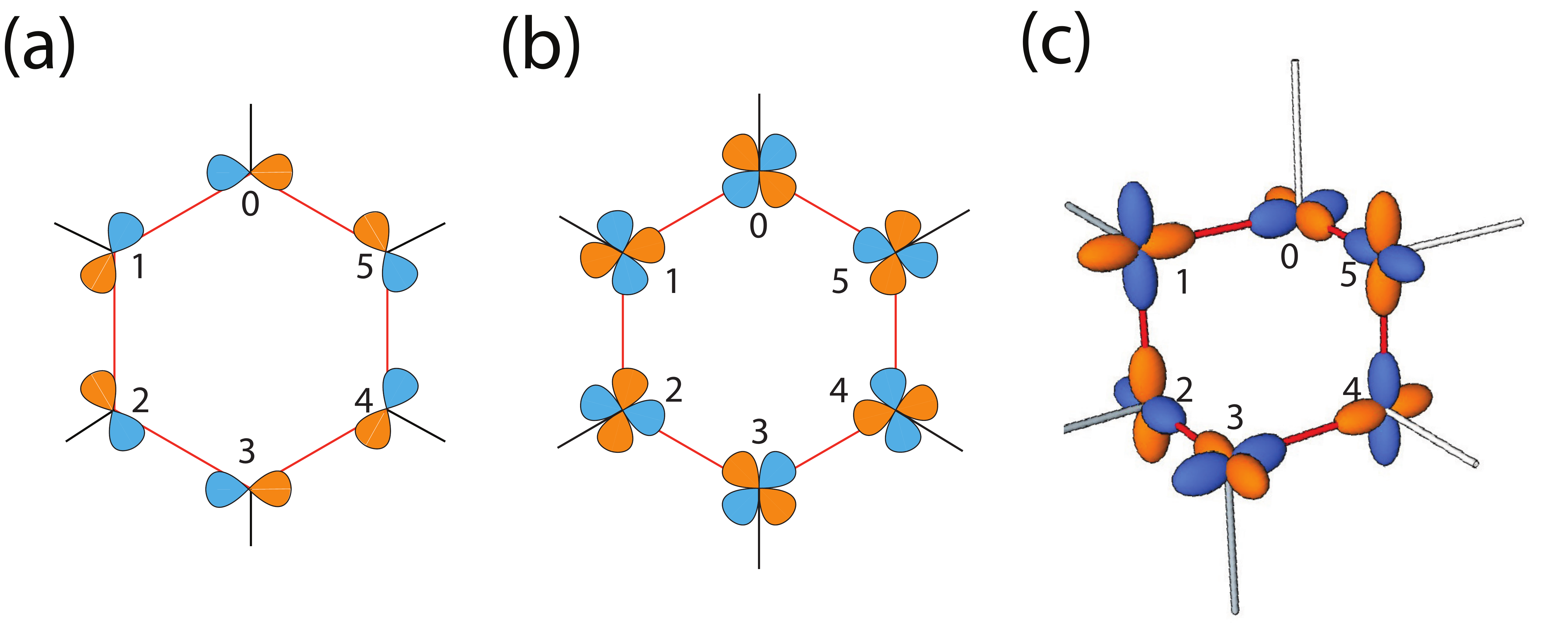}
\includegraphics[height=0.6\columnwidth]
{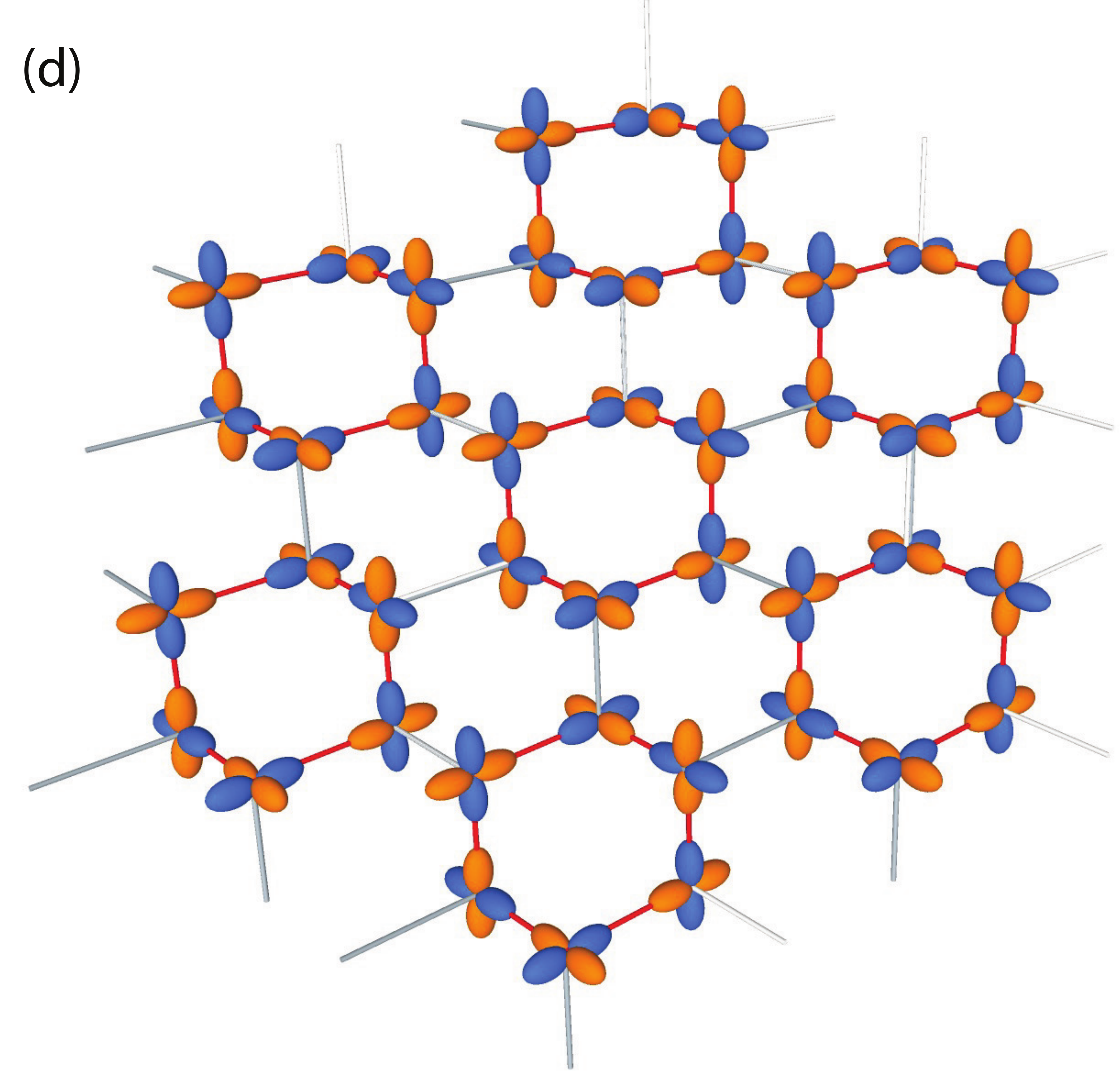}
\caption{(a) The spatial localized states in the lower flat band of the tight-binding Hamiltonian Eq.~\eqref{eq:tight-binding} for (a) $(p_x, p_y)$ doublet; (b) $(d_{xz}, d_{yz})$ doublet; (c) $E_g$ doublet. (d) The spatial localized state Wigner-crystal
when the lower flat-band is $1/3$ filled.}
\label{fig:flat_band}
\end{figure}

\subsection{The appearance of the flat-band and the localized state}
The existence of the flat bands implies that one can construct 
local eigenstates of the single-particle Hamiltonian. 
The flat band has been studied in detail in~\cite{Wu2007,Wu2008}  
in the context of the $p$-orbitals in the honeycomb optical lattice, 
and plaquette states on a hexagon are constructed as the local basis (Fig.~\ref{fig:flat_band}(a)).
Here we investigate it in the orbital-active Dirac material 
realized by the $\dxy$ and the $e_g$ doublets.

The localized states can be elegantly constructed from the Bloch wavefunction in Eq.~\eqref{eq:psi_14},
\bea
\ket{\psi_{\vec R}}_{\pm}=\frac{1}{\sqrt{N_k}} \sum \limits_k\ket{\psi(\vec k)}_{1,4} e^{-i \vec k \cdot  \vec R},
\eea
where $\vec R$ are the centers of the hexagons.  Each hexagon hosts one localized state from each flat band. 
The localized states are
\begin{widetext}
\bea
\label{eq:local}
\ket{\psi_{\vec R}}_{\pm}=
  \sum\limits_{n=0}^5(\pm1)^{n}  \left (\cos\frac{n\pi}{3}\gamma_x^\dagger(\vec{r}_n)+\sin \frac{n\pi}{3}\gamma_y^\dagger(\vec{r}_n)\right )\ket{0}.
\eea
\end{widetext}

The summation is over the six vertices of the hexagon as shown in Fig.~\ref{fig:flat_band}(b) for the case of the $\dxy$ doublet.  The localized state has the same weight on each site but different orbital configurations, related by $\frac{\pi}{3}$ rotations. On each site, the orbital is projected into the $\pi$-bonding along the outward bond away from the hexagon.  Due to the destructive interference, electrons in such localized single-particle states cannot leak out the plaquette,  rendering the localized states eigenstates of the tight-binding Hamiltonian in Eq.~\eqref{eq:tight-binding} with the same energy.
The above analysis can be carried over to the $E_g$ doublets on the buckled honeycomb lattice. In this case, the localized state is confined in a buckled hexagon as shown in Fig.~\ref{fig:flat_band}$(c)$.

\subsection{Orbital configurations at high symmetric points}

As shown in Fig.~\ref{fig:spectrum}($b$), the spectrum exhibits 
double degeneracy at the $K(K')$ point and the $\Gamma$ point 
in the Brillouin zone. 
Now we investigate the Bloch wavefunction at these high symmetric
points in detail.

\subsubsection{$\Gamma$ point}
Around the center of the Brillouin zone $\vec k=(0,0)$, the Hamiltonian Eq.~\eqref{eq:H_k}, in the unit of $t_\parallel$,  can be expanded as
\begin{widetext}
\bea
H_\Gamma (k)
=\frac{3}{2}  (1-\frac{1}{4} |k|^2) \tau_1 \otimes \sigma_0 - \frac{3}{4}  \left (  k_x \tau_2 \otimes \sigma_1 +k_y \tau_2  \otimes \sigma_3 \right ) 
-\frac{3}{16}  \left ( (k_x^2-k_y^2)\tau_1 \otimes \sigma_3+ 2 k_x k_y \tau_1 \otimes \sigma_1 \right ),
\label{H_Gamma}
\eea
\end{widetext}
where the Pauli matrices $\sigma_0 \sim \sigma_3$ ($\sigma_0$ represents the identity matrix) describe  the orbital degrees of freedom $\gamma_{x,y}$ in the $E$ irrep, and $\tau_0 \sim \tau_3$ act on the space of sublattice $(A, B)$.

To the leading order, the dispersion of the above band structure 
is
\bea
&E^\Gamma_{1,4}=\mp \frac{3}{2} t_\parallel \\
&E^\Gamma_{2,3}=\pm\frac{3}{2}t_\parallel \left  ( -1+\frac{1}{4}|\Delta k|^2 \right ).
\eea
Therefore, the bands touch each other quadratically at both upper and lower degeneracy points.
The degenerate wavefunctions at each touching point can be regrouped so that they only contain one of each the orbital component. At the lower touching point, the wavefunctions are
\bea
\ket{\psi(\Gamma)}^+_{x(y)}=\frac{1}{\sqrt{2}}\left ( \gamma_{x(y),A}^\dagger + \gamma_{x(y),B}^\dagger \right )\ket{0}.
\label{eq:psi_Gamma_1}
\eea
At the upper touching point, the $B$ sublattice component acquires a minus sign, and the wavefunctions are
\bea
\ket{\psi(\Gamma)}^-_{x(y)}=\frac{1}{\sqrt{2}}\left ( \gamma_{x(y),A}^\dagger - \gamma_{x(y),B}^\dagger \right )\ket{0}.
\label{eq:psi_Gamma_2}
\eea

\subsubsection{$K$ and $K'$ points}
Around $ \vec K=(\frac{4\pi}{3\sqrt{3}},0)$, the Hamiltonian in Eq.~\eqref{eq:H_k} can be expanded as
\bea
H_K =&-\frac{4}{3}\Delta k_x \tau_1 \otimes \sigma_0 + \frac{4}{3}\Delta k_y \tau_2 \otimes \sigma_0 \\
&-\frac{3}{8}(2+\Delta k_x) \tau_1 \otimes \sigma_0 - \frac{3}{8}\Delta k_y  \tau_2 \otimes \sigma_3 \\
&-\frac{3}{8} \Delta k_y \tau_1 \otimes \sigma_1- \frac{3}{8} (2-\Delta k_x)\tau_2 \otimes \sigma_1
\label{eq:H_K},
\eea
where $\Delta \vec k =\vec k- \vec K$.
The middle two bands touch each other with the dispersion,
\bea
E_{2,3}= \mp \frac{3}{4}t_\parallel |\Delta k|,
\eea
which demonstrates the Dirac cone.
The doubly degenerate wavefunctions can be combined so that each of them only occupies one of the sublattices:
\bea
\ket{\psi ( \vec K)}_A=\frac{1}{\sqrt 2}\left ( \gamma^\dagger _{x,A}+i \gamma^\dagger_{y,A} \right ) \ket{0}, \\
\ket{\psi (\vec K)}_B=\frac{1}{\sqrt 2}\left ( \gamma^\dagger _{x,B}-i \gamma^\dagger_{y,B} \right ) \ket{0}.
\label{eq:psi_K}
\eea

The orbital states in Eq. (\ref{eq:psi_K}) on the two sublattices 
are circularly polarized and exhibit opposite chiralities. 
Such complex combinations of orbitals exhibit distinct 
physical properties for different orbital realizations. 
In the case of the $(p_x,p_y)$ doublet as well as the $(d_{xz}, d_{yz})$ doublet, the circularly polarized state $\ket{\gamma_1}\pm i \ket{\gamma_2}$ carries angular momentum $L_z=\pm 1$; in the case of the $\dxy$ doublet, the circularly polarized state carries angular momentum $\mp 2$, which are
equivalent to $\pm 1$ due the 3-fold rotation symmetry; 
in the case of the $E_g$ doublet, it carries magnetic octupole moment. These complex orbital states play an important role in the topological properties of the orbital-active Dirac material and will be addressed in Sec.~\ref{sec:gap_opening}.

\subsection{Response of the flat band to magnetic field}
One interesting question regarding the flat band that appeared is its response to an external magnetic field. In a flat band, because the kinetic energy of the electrons is completely quenched, the usual semi-classical picture is no longer valid. Recent work~\cite{rhim2020flat} demonstrates that the response of flat bands to an external magnetic field is closely related to the quantum distance of the flat band. The quantum distance between two Bloch wavefunctions is defined as,
\bea
d=1-|\bra{\psi(k)}\psi(k')\rangle|^2,
\eea
which ranges from 0 to 1. The flat band is singular if $d$ is nonzero in the limit that $|k-k'|\rightarrow 0$. A singular point $k_0$ can be characterized by the maximal quantum distance $d_{max}$ between the wavefunctions of $k$ and $k'$ that are sufficiently close to $k_0$. In systems without orbital degrees of freedom, it is found that when $d_{max}$ is nonzero, the flat band splits into Landau levels in an energy window, and the width of the energy window is determined by $d_{max}$.

In our case, the flat band touches the dispersive Dirac band at the $\Gamma$ point.
The wavefunction of the flat band, expanded around the $\Gamma$ point, is
\bea
\ket{\psi}=\frac{1}{\sqrt{2}}\left (\sin \theta(\vec k), - \cos \theta(\vec k), \sin \theta(\vec k), -\cos\theta(\vec k) \right)
\label{eq:singular}
\eea
where $\theta(\vec k)$ is the azimuth angle of $\vec{k}$.
Therefore, the wavefunction at $\theta$ and $\theta-\pi/2$ are orthogonal to each other,  rendering the maximal quantum distance $d=1$. As a result, the flat band that appeared here is singular by definition.

\begin{figure}
\includegraphics[height=0.49\columnwidth, width=0.49\columnwidth]
{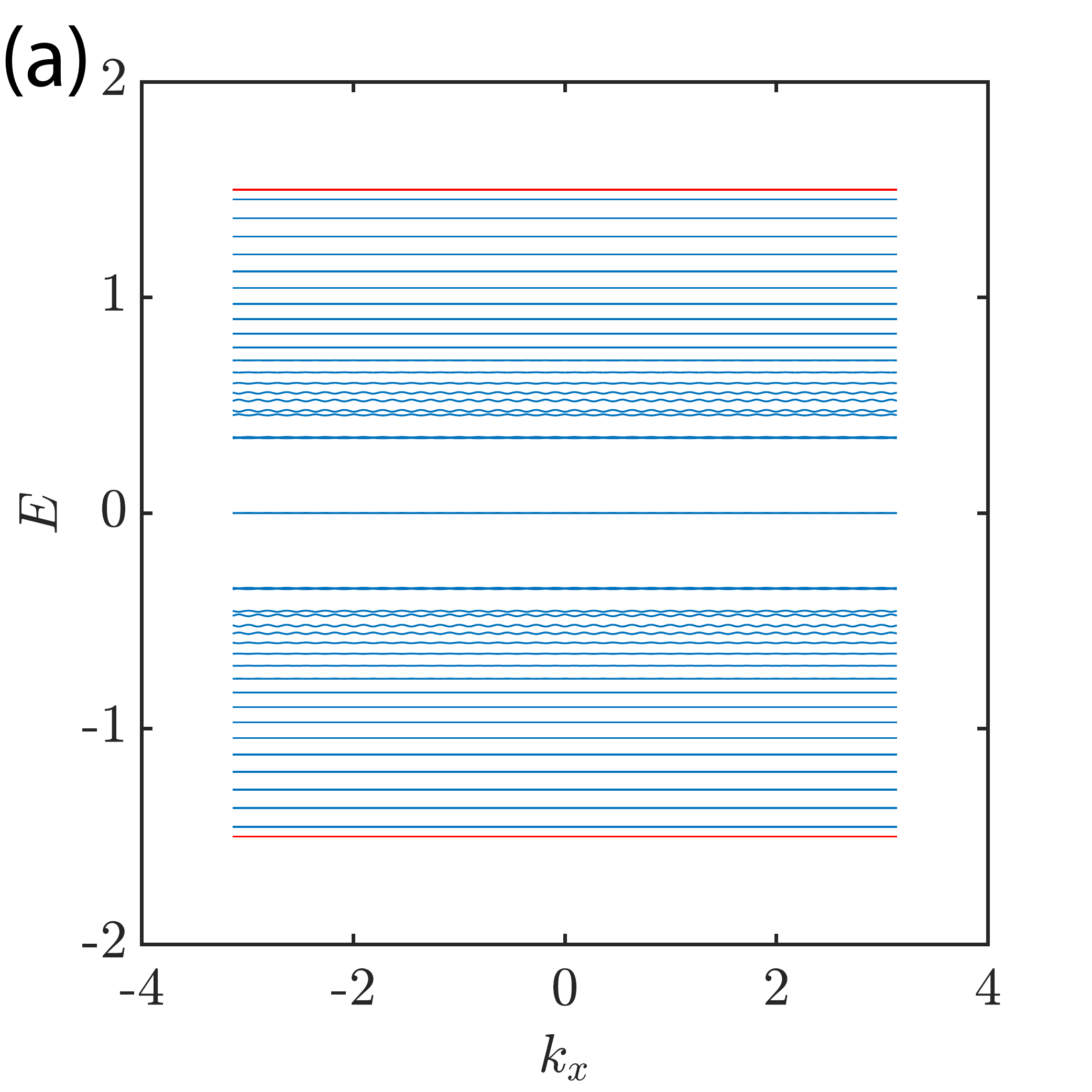}
\includegraphics[height=0.49\columnwidth, width=0.49\columnwidth]
{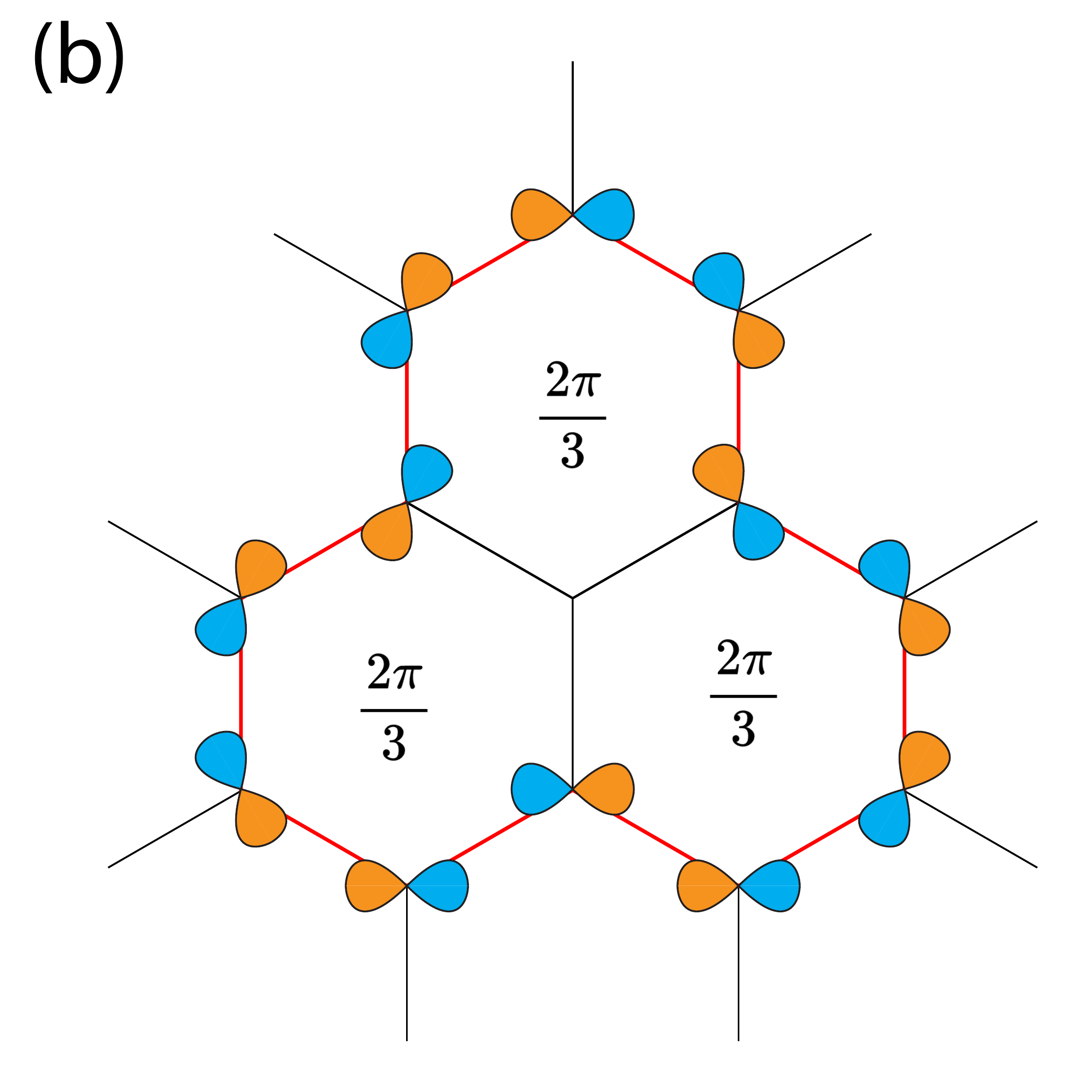}
\caption{(a) The flat band in the orbital active Dirac system is singular but does not respond to an external magnetic field due to the orbital nature of the singularity. (b) An example of the localized state in the presence of the magnetic field. Here the flux through each hexagon is $2\pi/3$. The localized state occupies three hexagons.}
\label{fig:magnetic_field}
\end{figure}

We study the response of the flat band to an external magnetic field by including the magnetic field in the hopping parameter 
$t\rightarrow t\exp(i\int A(\vec r)d\vec r)$. 
For simplicity, the Landau gauge $A=B(0,x)$ is chosen and the strength 
of the magnetic field $B$ is set by the flux through each hexagon $2\pi p/q$. 
In the presence of the magnetic field, the original four bands split 
into $4q$ sub-bands as shown in Fig.~\ref{fig:magnetic_field}. 
While the middle two dispersive Dirac bands form the characteristic 
Landau levels, surprisingly, the two singular flat bands are completely
inert to the magnetic field, in contrast with previous results on 
singular flat bands without orbital degrees of freedom.

The reason is due to the orbital nature of the singularity in Eq.~\eqref{eq:singular}. 
In the presence of the magnetic field, one can still construct 
localized states inside the flat band. 
Instead of occupying a hexagon, the localized states now 
occupy a magnetic unit cell. 
The wavefunction is nonzero only along the boundary of the
magnetic unit cell, and the orbital configuration is 
parallel to the tangential direction.
An example of the localized states is shown in Fig.~\ref{fig:magnetic_field}(b) for the 
flux given by $\frac{2\pi}{3}$.

\section{General symmetry consideration beyond strong 
anisotropic limit}
\label{sec:general_symmetry}

In the last section, we demonstrate many remarkable features 
resulting from the tight-binding Hamiltonian Eq.~\eqref{eq:tight-binding},
including orbital enriched Dirac cone, quadratic band touching 
and flat bands. 
Since the Hamiltonian only includes the nearest neighboring
$\sigma$-bonding, a natural question is whether these features 
rely on the specific form of the Hamiltonian, or, 
are protected by symmetry.

In this section, we address this issue by general symmetry consideration.
We study the band structure of the orbital active Dirac materials 
around the high symmetric points in the Brillouin zone using 
$k\cdot p$ theory.
In general, the band flatness is not generic and can be bent by the $\pi$-bondings.  
However, the orbital configurations at the high symmetric points 
and $k$ dependence of the dispersion around the $\Gamma$ and $K$ 
points are preserved as long as the symmetry of the system is 
respected. 
In the following, we consider the effects of the point group symmetry 
of the buckled honeycomb lattice $D_{3d}$.

\subsection{$\Gamma$ point}
At the $\Gamma$ point, the group of wavevector is the point group 
of the lattice, $D_{3d}$, which has an inversion symmetry.
It has 6 irreps: $A_{1g}$, $A_{2g}$, $A_{1u}$, $A_{2u}$,
$E_g$, and $E_u$.
The Bloch wavefunction is composed of the orbital part 
$\ket{\gamma_{x,y}}$ and the plane wave part $\ket{\Gamma}_{A, B}$, 
which can be classified into the irreps of the group of wavevector 
separately.
The orbital degrees of freedom $(\ket{\gamma_x}, \ket{\gamma_y})$ 
form the two dimensional $E_g$ irrep, while the plane wave part $\ket{\Gamma}_A + \ket{\Gamma}_B$ and $\ket{\Gamma}_A - \ket{\Gamma}_B$ 
form the 1D $A_{1g}$ and $A_{2u}$ irreps.
Therefore the composite Bloch wavefunction can be grouped into two two-dimensional irreps, $E_g$ and $E_u$, respectively.

This indicates that four energy levels can be grouped into two doubly-degenerate sets, where the degeneracy completely originates 
from orbitals.  
The Bloch wavefunctions of the $E_g$ irrep are
\bea
\ket{\gamma_{x(y)}}\otimes (\ket{\Gamma}_A+\ket{\Gamma}_B),
\eea
and the wavefunctions of the $E_u$ irrep are
\bea
\ket{\gamma_{x(y)}}\otimes (\ket{\Gamma}_A-\ket{\Gamma}_B),
\eea
This is consistent with Eq.~\eqref{eq:psi_Gamma_1} and Eq.~\eqref{eq:psi_Gamma_2}. Furthermore, in both degeneracy sets, the orbital part of the wavefunctions  can be regrouped into the complex orbital states $\ket{\gamma_{\pm}}=\ket{\gamma_1} \pm i\ket{\gamma_2}$. Therefore, spin-orbit coupling is able to gap out the degeneracy.

The generic dispersion around the $\Gamma$ point can be obtained from the $k \cdot p$ theory. The $k \cdot p$ Hamiltonian is invariant under the $D_{3d}$ point group symmetry.  
In order to write down all the symmetry allowed terms, 
it is convenient to first classify the operators in the 
orbital space $\vec \sigma$, the operators in the sublattice 
space $\vec \tau$  and the momentum $\vec k$ into irreps 
of the point group $D_{3d}$.
In the sublattice space, the classification is
\bea
\begin{cases}
\tau_0,  \tau_1  & A_{1_g} \\
\tau_2, \tau_3   & A_{2_u}.
\end{cases}
\eea
In the orbital space, the classification is
\bea
\begin{cases}
\sigma_0 , & A_{1g} \\
\sigma_2, & A_{2g} \\
(\sigma_1, \sigma_3), & E_g.
\end{cases}
\eea
For the momentum, the classification is
 \bea
\begin{cases}
(k_x, k_y) & E_{u}\\
k_x^2+k_y^2 & A_{1g} \\
(2k_x k_y, k_x^2-k_y^2), & E_{g}.
\end{cases}
\eea

Based on these classifications and the product table of the $D_{3d}$ 
point group, the  most general Hamiltonian takes the following form,
\begin{widetext}
\bea
H_\Gamma (\vec k)=&(h_1+h_2 ( k_x^2+ k_y^2))\tau_1 \otimes \sigma_0 +
 k_x (h_3  \tau_2 \otimes \sigma_1 + h_4  \tau_3 \otimes \sigma_1 )+  k_y (h_3 \tau_2 \otimes \sigma_3  + h_4  \tau_3 \otimes \sigma_3 )+\\
&(k_x^2- k_y^2)(h_5 \tau_0 \otimes  \sigma_3 + h_6 \tau_1 \otimes \sigma_3 )+
2 k_x  k_y (h_5 \tau_0 \otimes \sigma_1 +h_6 \tau_1 \otimes \sigma_1 ),
\label{H_Gamma_general}
\eea
\end{widetext}
where $h_1\sim h_6$ are constants with the unit of energy. 
At the first  order of $\vec  k$, the degeneracy is still preserved, 
so we have to include second-order terms of $\vec k$.
It recovers the tight-binding Hamiltonian around the $\Gamma$ point 
presented in Eq.~\eqref{H_Gamma}, when the $h$'s are set to
\bea
&h_1=-4 h_2=-2h_3 = -8 h_6=\frac{3}{2}t_\parallel \\
&h_4=h_5=0,
\eea

In the leading order, the dispersion is
\bea
&E_{1,2}&=-h_1-\left (2h_2+\frac{h_3^2+h_4^2}{h_1}\right )| k|^2\pm (h_5-h_6)| k|^2\\
&E_{3,4}&=h_1+\left (2h_2+\frac{h_3^2+h_4^2}{h_1}\right )|k|^2\pm (h_5+h_6)| k|^2.
\eea
At finite $k$, the degeneracy is lifted by
\bea
&|E^\Gamma_1 - E^\Gamma_2 |=\frac{m_-}{2}|k|^2 \\
&|E^\Gamma_3 - E^\Gamma_4 |=\frac{m_+}{2} |k|^2 . \\
\eea
where the effect mass $m_\pm=4 |h_5 \pm h_6| $.  Therefore, the bands touch quadratically at both degenerate points. It is known that quadratic bound touching is unstable to interaction and can lead to exotic phases such quantum Hall effect and nematicity~\cite{Sun2011}.

\subsection{$K(K')$ point}
At the $K$ point, the group of wavevector is $D_3$, containing the 
three-fold rotations around the perpendicular axis and three
2-fold rotations around horizontal axis that interchanges 
the two sublattices.

The Bloch wavefunctions $\ket{\psi(K)}$, containing both the plane wave part and the orbital part, can be organized into irreps of the group of 
wavevector. 
The plane wave part contains two sublattice components, forming 
the $E$-irrep, with the $A/B$ sublattice component carrying 
chirality $\pm1$. 
The on-site orbital degrees of freedom $\gamma_x$ and $\gamma_y$ 
also transform as the $E$ irrep, the complex combination 
$\gamma_x\pm i \gamma_y$ carrying the chirality $\pm 1$.
Therefore, the four composite wavefunctions can be decomposed into
three irreps as $2\otimes2=1\oplus1\oplus 2$.
There are two trivial $A_1$ representations, where the chiralities of
the orbital and planewave cancel each other and an $E$ irrep where
the chiralities of the orbital and planewave add up.
In general, the two $A_1$ states do not have the same energy. 
In contrast, the two states in the $E$ irrep are degenerate from 
symmetry and carry opposite chirality at the Dirac point.
Explicitly, the two states are:
\bea
&\left(\ket{\gamma_x}+ i\ket{\gamma_y}\right )\otimes \ket{K}_A\\
&\left(\ket{\gamma_x}- i\ket{\gamma_y}\right)\otimes \ket{K}_B.
\label{eq:wave_k}
\eea
This is consistent with the wavefunctions of the tight-binding model at $K$ in Eq.~\eqref{eq:psi_K}.

To obtain the generic dispersion around $K(K')$, we again employ the $k\cdot p$ theory.
The Hamiltonian around $K$,  a combination of the plane wave, orbital, and sublattice has to be invariant under the $C_{3v}$ point group. Following the same strategy, we  first organize $\vec{\sigma}$,  $\vec{\tau}$ and the momentum $\Delta \vec k =\vec k-\vec K$ into irreps of the little group $d_3$.
In the sublattice space, we have,
\bea
\begin{cases}
\tau_0, &A_1 \\
\tau_3, &A_2 \\
(\tau_1, -\tau_2), &E.
\end{cases}
\eea
In orbital space, we have,
\bea
\begin{cases}
\sigma_0 , &A_1 \\
\sigma_2,  &A_2 \\
(\sigma_3, -\sigma_1), &E.
\end{cases}
\eea
In addition, the momentum $(\Delta k_x, \Delta k_y)$ belongs to the $E$ irrep as well.

Therefore, based on the product table of $D_3$ point group, the most general Hamiltonian, apart from  an overall constant, reads,
\begin{widetext}
\bea
H_K(\Delta \vec k)=&h_1  \tau_3  \otimes  \sigma_2 + h_2 ( \tau_1 \otimes  \sigma_3 + \tau_2 \otimes \sigma_1)+\\
&\Delta k_x \left \{+ h_3 \tau_0 \otimes \sigma_3+h_4 \tau_1 \otimes \sigma_0 +  h_5  \tau_2 \otimes \sigma_2 + h_6 \tau_3 \otimes \sigma_1 +h_7 ( \tau_1 \otimes  \sigma_3 - \tau_2 \otimes \sigma_1)\right \}+\\
&\Delta k_y \left \{ -h_3\tau_0 \otimes \sigma_1 - h_4 \tau_2 \otimes \sigma_0 + h_5 \tau_1 \otimes \sigma_2 + h_6  \tau_3 \otimes \sigma_3 +  h_7 ( \tau_2 \otimes \sigma_3 + \tau_1 \otimes \sigma_1)\right \}.
\label{eq: H_K_general}
\eea
\end{widetext}
The expansion of the $\sigma$-bonding Hamiltonian at $K$ 
in Eq.~\eqref{eq:H_K} is a  special case with
\bea
h_2=h_4=2 h_7=-\frac{3}{4} t_\parallel \\
h_1=h_3=h_5=h_6=0.
\eea
In the general situation, at the leading order of $\Delta \vec k$, the dispersions of the four bands read,
\bea
E^K_{1,4}&=-h_1\pm 2 h_2+ \mathcal{O}(|\Delta k|^2)  \\
E^K_{2,3}&=h1 \pm 2 h_7 |\Delta k| +\mathcal{O}(|\Delta k|^2),
\eea
which is consistent with those given by the nearest-neighboring  tight-binding model. The dispersion $E^K_{2,3}$ is Dirac-like as long as $h_7 \neq 0$.
The situation of the $K'$ point can be obtained by performing the reflection symmetry with respect to the $y$ axis.

The above analysis solely relies on the non-Abelian nature of the 
point group and therefore is widely applicable to the 
orbital-active Dirac material, 
independent of the origin of the orbitals.

\section{Gap opening mechanism}
\label{sec:gap_opening}
We have shown that the symmetry of the honeycomb lattice protects the degeneracy of the band structure at $K(K')$ point and $\Gamma$ point. The degenerate states form the 2-dimensional irrep of the little group at the high symmetry points.
The degeneracy can be lifted by including various symmetry-breaking terms in the Hamiltonian, which introduces gaps at the Dirac point or/and the quadratic band touching point.
The interplay of different symmetry-breaking terms can give rise to various topological band structures,  rendering the orbital active Dirac system a flexible platform for realizing topological insulators with different edge-state properties. In this section, we discuss the gap opening mechanisms for different orbital doublets of the $E$ irrep, previously studied in different contexts~\cite{Zhang2014,Xiao2011}, in a unified manner.


Based on Eq.~\eqref{eq:psi_Gamma_1} and Eq.~\eqref{eq:psi_K}, the degenerate wavefunctions at $\Gamma$ and $K$ can be grouped into circular polarized orbital state $\gamma_x \pm i\gamma_y$ with opposite chirality. As a result, a $\sigma_2$ term in the orbital space, which measures the chirality, is able to lift the degeneracy at both $\Gamma$ and $K$ points.
In addition, at $K(K')$ points, the two complex orbital states only occupy A and B sublattices, respectively. As a result, a $\tau_3$ term in the sublattice space can also gap out the Dirac points. In contrast, since the degenerate states at the $\Gamma$ point have the same weight on both sublattices, they remain degenerate after the $\tau_3$ term is added to the Hamiltonian. One can also add other terms to the $k\cdot p$ Hamiltonian to open up a gap in the spectrum. But the two terms mentioned above, $\tau_3$ and $\sigma_2$, denoted as $H_m$ and $H_\lambda$ respectively, are among the simplest and have a clear physical origin. In the real space, they have the following form,
\bea
H_m&=m\left\{\sum\limits_{\vec r\in A ,\sigma} \gamma^\dagger_{\sigma}(\vec r) \gamma_{\sigma}(\vec r)-\sum\limits_{\vec r\in B ,\tau} \gamma^\dagger_{\sigma}(\vec r) \gamma_{\sigma}(\vec r)\right\}\\
H_\lambda&=\lambda\left\{  \sum\limits_ {\vec r \in A, B} i \gamma_x^\dagger (\vec r) \gamma_y (\vec r ) + h.c. \right \}.
\eea
The term  $H_m$ represents the staggering mass resulting from the imbalance between the $A$ and $B$ sublattices, which for example, occurs in TMD materials. It only depends on the particle number on each sublattice and does not rely on the particular orbital state the electrons occupy.
On the other hand, $H_\lambda$ measures the chirality of the orbital and originates from spin-orbit coupling $-\lambda_0\vec{L}\cdot \vec{S}$, where $\lambda_0$ is the atomic spin-orbit coupling strength,  $\vec{L}$ is the physical angular momentum operator and $\vec{S}$ is the spin operator of electrons.

In free space, $\vec{L}$ acts on the Hilbert space labeled by the angular momentum $s$, $p$, $d$, etc. In the case of the planar and buckled honeycomb lattices, the spherical symmetry reduces to the site symmetry $C_{3v}$. As the result, the physical angular momentum $\vec{L}$ should be projected into the 2d irrep.
The result depends on the particular orbital realizations of the irrep, even though they are equivalent under the $C_{3v}$ point group.
In the following, we discuss the different orbital realizations case by case.

\begin{figure*}
\includegraphics[height=0.32\textwidth, width=0.32\textwidth]
{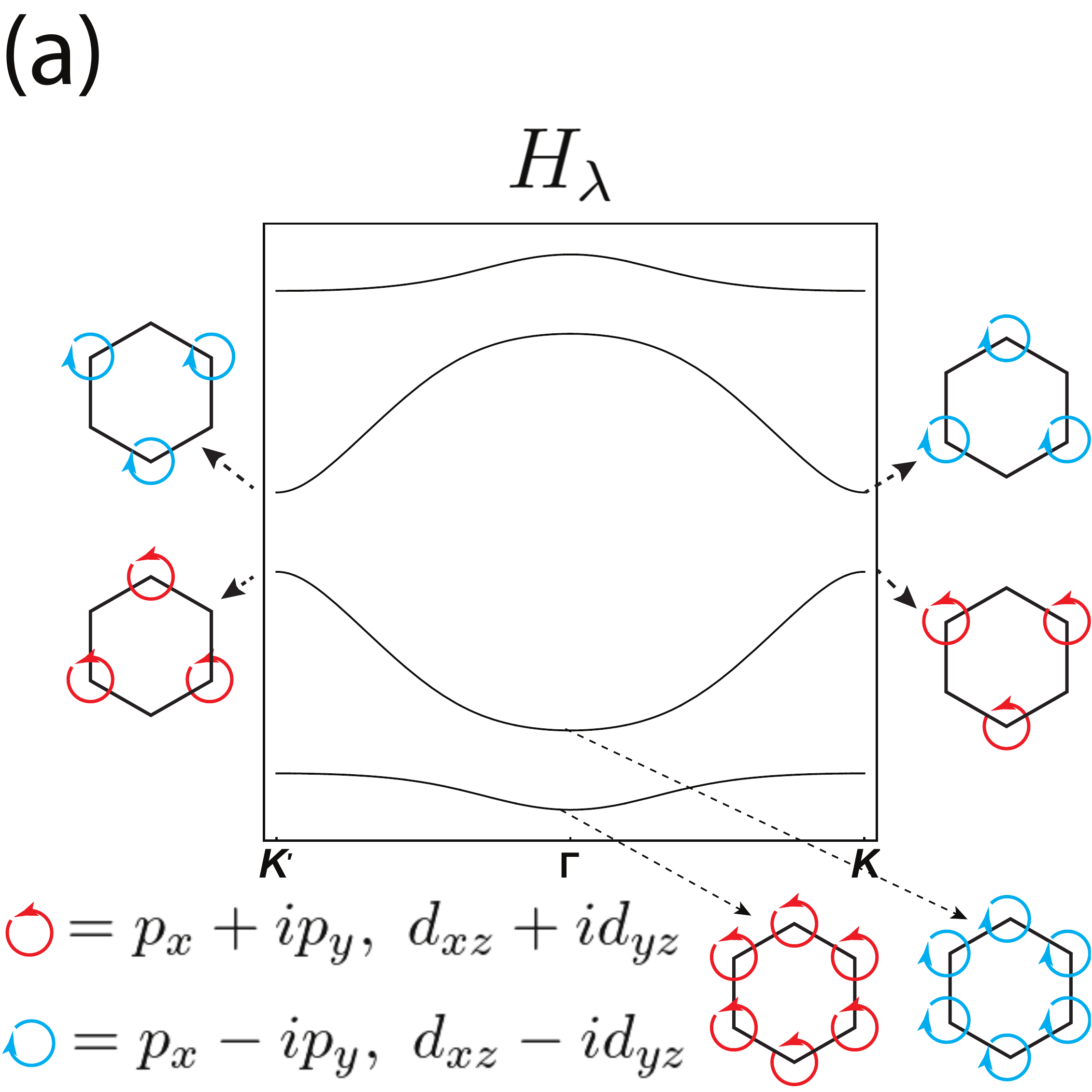}
\includegraphics[height=0.32\textwidth, width=0.32\textwidth]
{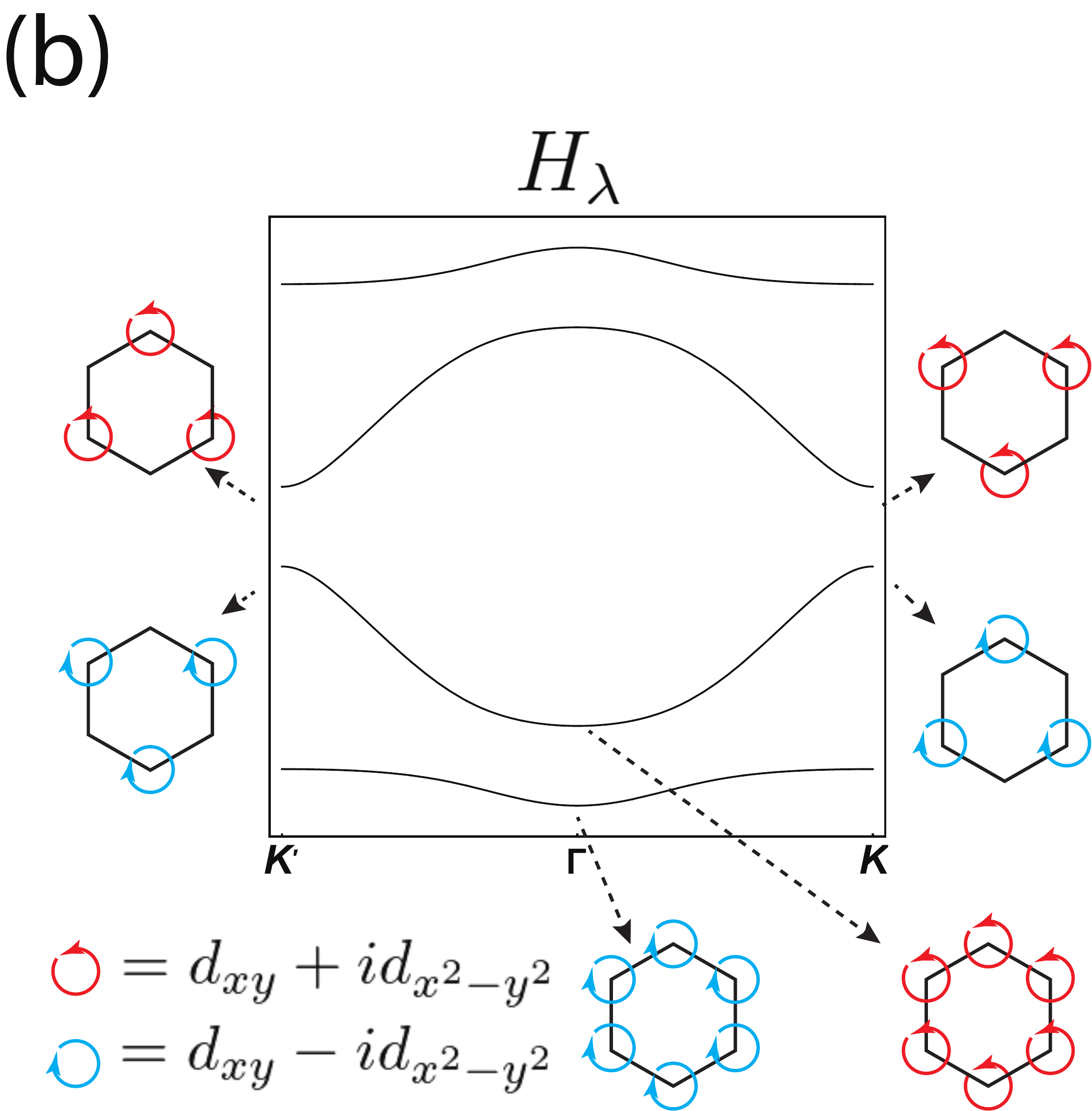}
\includegraphics[height=0.32\textwidth, width=0.32\textwidth]
{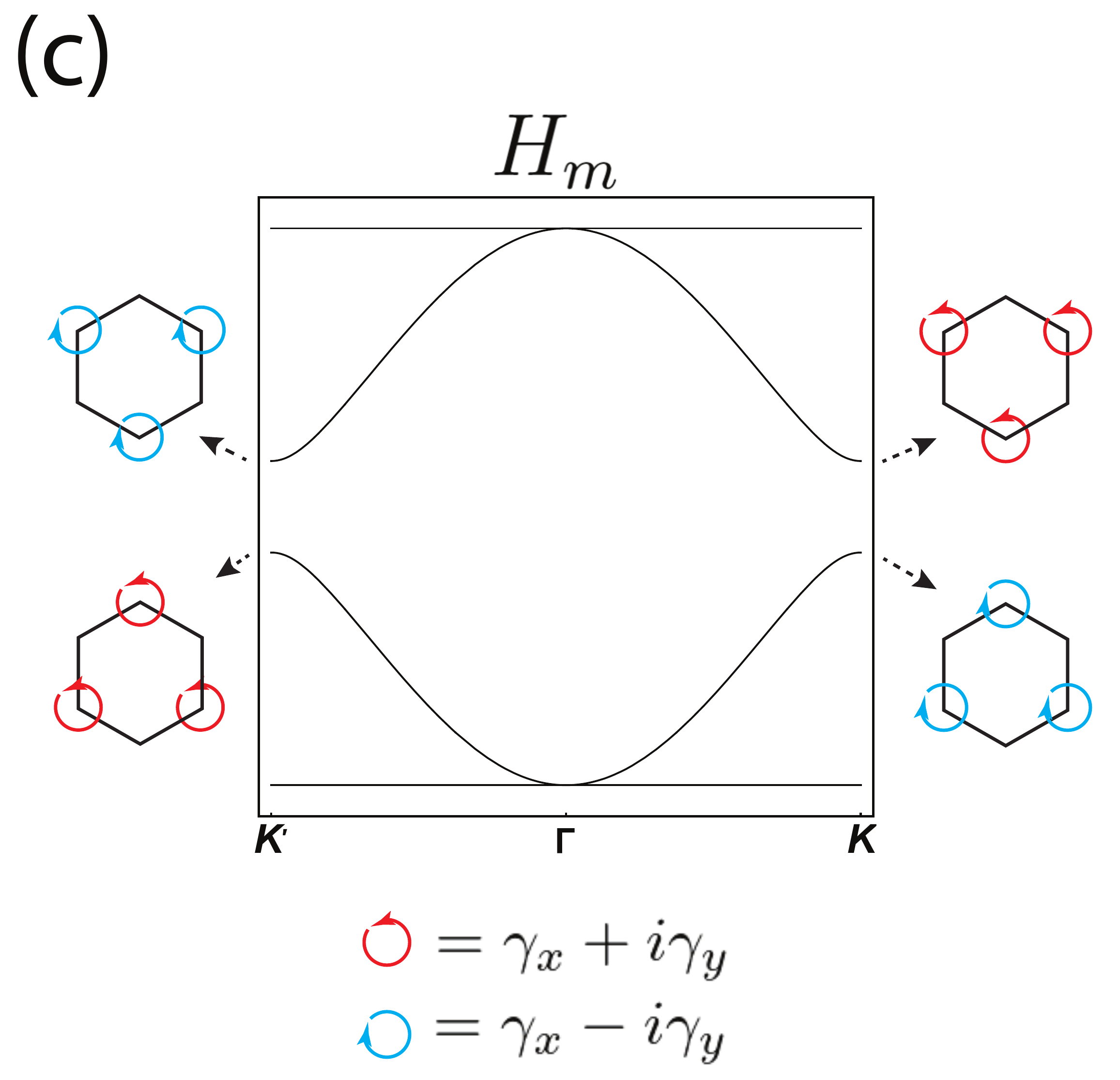}
\caption{(a) The gap opening pattern from the $H_\lambda$ term for the $(p_x, p_y)$ and $(d_{xz},d_{yz})$ doublets. (b) The gap opening pattern from the $H_\lambda$ term for the $\dxy$ doublet. (c) The gap opening pattern from the stagger mass term for all cases. The complex orbitals state naturally occurs at the $K (K')$ point. }
\label{fig:gap}
\end{figure*}

In the case of the $(p_x, p_y)$ and $(d_{xz}, d_{yz})$ doublets, the circular polarized orbital state have angular momentum $\pm 1$. The $L_z$ operator, projecting into the two-dimensional space, becomes $\sigma_2$, while $L_x$ and $L_y$ are zero. The spin orbit coupling $-\lambda \vec{L}\cdot \vec{s}$ becomes $-\lambda \sigma_y s_z$. As a result, after the spin-orbit coupling is included in the Hamiltonian, which lifts the degeneracy at the $\Gamma$ point and the $K$ points, the complex orbital state with positive chirality, $p_x + i p_y$ or $d_{xz} + i d_{yz}$, has higher energy than its partner, as shown in Fig.~\ref{fig:gap}(a).
In the case of the $(d_{xy}, d_{x^2-y^2})$ doublet, the complex orbital states carry angular momentum $\mp 2$. The $L_z$ operator in this space is $-2\sigma_y$ while the other components vanish. Therefore, the spin-orbit coupling term is $2\lambda_0 \sigma_y s_z$. Note the factor of 2 and extra minus sign compared with the previous two cases. Therefore, the complex orbital state $d_{xy} + i d_{x^2-y^2}$ has lower energy than its partner at the $\Gamma$ point and the $K$ points when the degeneracy is lifted, as shown in Fig.~\ref{fig:gap}(b).

The $E_g$ doublet is special. As discussed in Sec.~\ref{sec:octupole}, the complex orbital combinations do not carry angular momentum. The angular momentum operator $L$, projecting into this space, vanishes for all components. The $\sigma_y$ term measure the octupole momentum $\hat{f}_{xyz}$ instead, which is the lowest rank of non-vanishing multipole order for $E_g$ orbitals.
On the level of single-particle physics, the $\sigma_y$ term cannot be obtained directly from the spin-orbit coupling in the $E_g$ space. It can appear as a result of second-order perturbation, taking into account the virtual excitation from the $E_g$ orbitals to the $t_{2g}$ orbitals.
As discussed in Eq.~\eqref{eq:111_irrep}, the $T_{2g}$ orbitals splits into one 1d irrep $A_1$ and one 2d irrep $E$ of the site symmetry group $C_{3v}$, which in general have distinct onsite energies.
We denote the energy difference from the two irreps derived from the $T_{2g}$ orbitals to the $E_g$ orbitals as $\Delta_1$ and $\Delta_2$, respectively.
 The second order spin-orbit  coupling reads,
\bea
&H_{\lambda}'\\
&=-\lambda_0^2P_{E_g}\left\{\frac{\vec{L}\otimes\vec S P_{A_{1g}}\vec{L}\otimes\vec S}{\Delta_1}+   \frac{\vec{L}\otimes\vec S P_{E}\vec{L}\otimes\vec S}{\Delta_2}\right\} P_{E_g}\\
&=-(\frac{\lambda_0^2}{2\Delta_1}+\frac{\lambda_0^2}{\Delta_2})\sigma_0-\frac{\lambda_0^2}{\Delta_1\Delta_2}(\Delta_1-\Delta_2)    \sigma_2\otimes S_{(1,1,1)},
\label{eq:hlambda}
\eea
where $P_{E_g}$, $P_E$ and $P_{A_1}$ are the projection operators, and $S_{(1,1,1)}$ is the spin operator along the (1,1,1) direction. The first term is proportional to the identity operator and thus can be absorbed into the chemical potential. The second represents the effective spin-orbit coupling in the $E_g$ doublets with the spin-orbit coupling strength,
\bea
\lambda = \frac{\lambda_0^2}{\Delta_1\Delta_2}(\Delta_1-\Delta_2).
\eea
Recall that $\Delta_1-\Delta_2$ is the energy difference between the $A_1$ and $E$ irreps derived from the $T_{2g}$ orbital in Eq.~\eqref{eq:111_irrep}. Therefore, the energy splitting of the $T_{2g}$ triplet under $C_{3v}$ site symmetry is essential for nonzero $\lambda$.


\begin{figure}
\includegraphics[height=0.49\columnwidth, width=0.49\columnwidth]
{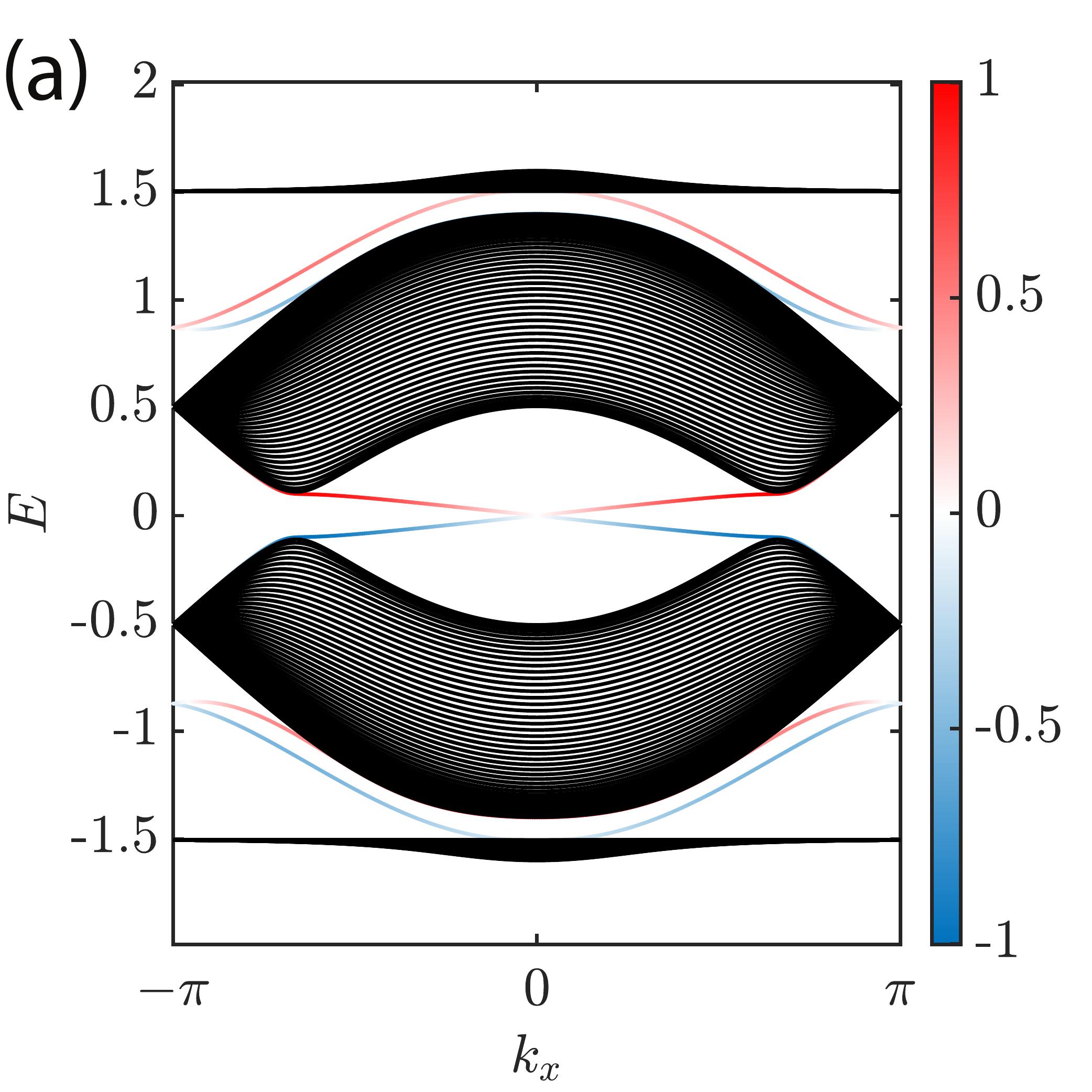}
\includegraphics[height=0.49\columnwidth, width=0.49\columnwidth]
{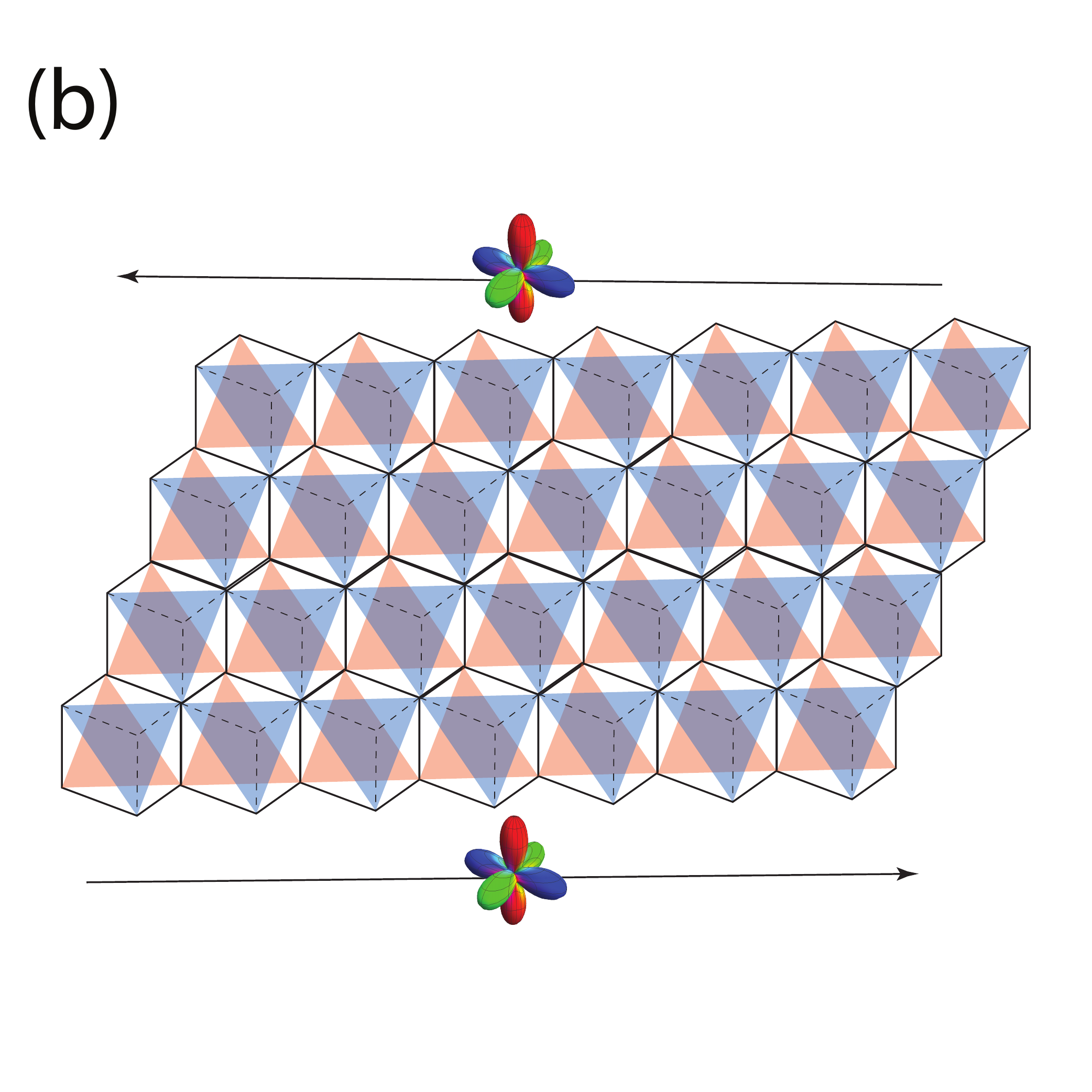}
\caption{(a) The spectrum of the Hamiltonian including $H_\lambda$ term on a ribbon geometry. $\lambda$ is set to $0.2t_\parallel$. Edge states appear between the bulk bands. The orbital wavefunction of the edges is complex, the chirality indicated by the color bar. (b) The ribbon geometry. When the orbital realization is from the $E_g$ doublet, the edge states carry magnetic octupole momentum. }
\label{fig:edge}
\end{figure}

The gaps introduced by the $\sigma_y$ term in the orbital space are topological. It is straightforward to show that the four bands in Fig.~\ref{fig:gap}(a) and (b) acquire Chern numbers 1, 0, 0, -1 from the bottom to the top. As a result, edge states appear on the boundary of the material. We consider the Hamiltonian on a ribbon with finite width but infinite length, in which case $k_x$ remains a good quantum number. The spectrum is plotted in Fig.~\ref{fig:edge}(a) as a function $k_x$, showing the edge states between the four bulk bands.
The orbital wavefunctions of the edge states are in general complex. The expectation value of the $\sigma_y$ operator in the orbital space is indicated by the color bar.
When the orbital degree of freedom is the $E_g$ doublet, the edge states carry the magnetic octupole moment instead of the dipole moment, sketched in Fig.~\ref{fig:edge}(b).

The topological gaps are proportional to the coefficient $\lambda$ of the $\sigma_2$ term  in the Hamiltonian. Remarkably, when the orbital degrees freedom are realized by the $(p_x, p_y)$, $(d_{xz}, d_{yz})$ or $(d_{xy}, d_{x^2-y^2})$, $\lambda$ is directly related to the atomic spin-orbit coupling strength $\lambda_0$, which could be quite large for heavy atoms. This leads to a robust topological phase, such quantum spin hall effect, at high temperatures.
In contrast, in $E_g$ systems, the $\sigma_2$ term comes from the second-order perturbation of the spin-orbit coupling. Therefore, in the $E_g$ Dirac materials, the band degeneracy is much more stable than other realizations. Even though the quadratic band touching can be gaped out from dynamic spin-orbit  coupling generated by interaction, the Dirac points are stable against interaction, making the $E_g$ Dirac materials an ideal 2D Dirac semi-metal.

We also present the gap opening pattern from the staggering mass term $H_m$ in Fig.~\ref{fig:gap}(c), which is the same for different orbital realizations. $H_m$ term leads to a trivial band insulator by itself
However, including $H_m$ in the presence of $H_\lambda$ can lead to richer topological phases with various Chern bands and edge state configurations~\cite{Zhang2014}.

\section{The interaction effects}
\label{sec:interaction}
The interplay between band structure and interactions can result in various interesting phases of matter, depending on the filling factors. Some of the most interesting phases are discussed below.

For simplicity, let us first consider the spinless fermions.
In this case, each site are maximally occupied by two fermions because of the orbital degrees of freedom.
At half-filling, i.e., one fermion per site,  the Fermi energy is at
the Dirac points, which are stable against weak interactions.
Therefore, the system remains a Dirac semi-metal for weak interactions
but becomes a Mott insulator for strong enough interactions.
Consider the simplest on-site interaction
\begin{equation}
\sum\limits_i V n_{x,i} n_{y,i}
\end{equation}
where $n_{x,i}$ and $n_{y,i}$ are the number operators for $p_x$ and
$p_y$ orbitals, respectively.
In the large $V$ limit, the system is expected to undergo a phase transition to a Mott insulating phase.
The orbital super-exchange is described by a quantum 120$^\circ$ model,
which is a frustrated orbital-exchange model.
The order-from-disorder analysis show the ground state possesses
a $\sqrt 3\times \sqrt 3$ type orbital-ordering~\cite{Wu2008a}.

At quarter filling, the Fermi level is at the quadratic band
touching point, which is unstable against interaction.
Infinitesimal interaction opens a gap, leading to the anomalous
quantum hall phase that breaks the time-reversal symmetry or
a nematic phase that breaks the rotational symmetry~\cite{Chen_kagome2018, Xiao2011, Ruegg2011, Ruegg2012}.

As the filling factor becomes smaller than 1/4, the Fermi level is within the flat band when the $\pi$-bonding is negelcted, and the system starts to develop different types of order orderings
at commensurate fillings~\cite{Wu2007}.
In particular, when the flat band is $1/3$-filled, the localized states close-pack the lattice.
If only on-site interactions are considered, such a close-packing
many-body state is the \textit{exact} many-body ground state.
This close-packing state breaks the original lattice translation
symmetry with an enlarged $\sqrt{3}\times \sqrt{3}$ unit cell,
sketched in Fig.~\ref{fig:flat_band}(b).
When long-range interactions are considered, the Wigner crystal
appears at even lower fillings.

Furthermore, when the Dirac band is 3/4-filled, the Fermi surface
is a regular hexagon by connecting the middle points
of the first Brillouin zone edge.
This causes Fermi surface nesting by three inequivalent momenta,
which makes the system unstable against weak interactions.
This can lead to the formation of exotic states of matter, such as
orbital density waves or superconductivity \cite{nandkishore2012chiral,wang2012functional}.	

Including the spin degrees of systems can further enrich the aforementioned phases. In this case, the on-site interaction takes the following form:
\begin{equation}
\begin{aligned}
H_{int} &=    U (n_{x}^\uparrow n_{x}^\downarrow + n_{x}^\uparrow n_{i,x}^\downarrow ) + \Delta \left(\gamma^{\uparrow,\dagger}_x \gamma^{\downarrow,\dagger}_x\gamma^{\downarrow}_y \gamma^{\uparrow}_y + h.c. \right) \\
    &- J (\vec S_x \cdot \vec S_y - \frac{1}{4} n_x n_y) + V n_xn_y
\end{aligned}
\end{equation}
where $U$ is the Hubbard interaction, $J$ is the Hund's coupling, $V$ is the inter-orbital repulsion and $\Delta$ is the pairing hopping term.

In the spinful case, at the half-filling, each site is occupied by two electrons.
Because of the strong intra-orbital repulsion and Hund's coupling, two electrons prefer to stay in two different orbitals and form a triplet.
As a result, the low-energy effective theory of the Mott insulator is described by a spin-1 Heisenberg model on the honeycomb lattice where
the orbital degrees of freedom are inert, which is in sharp contrast with the spinless model.

At the quarter-filling,  the bottom two spinful bands are filled. The Fermi surface is right at the quadratic band-touching point.
When the interaction is weak, it dynamically generates the spin-orbital coupling term, which gives rise to the quantum spin Hall effect.
As the interaction strength grows, the system becomes a Mott insulator where each site is occupied by one electron with both orbital and spin degrees of freedom, and the system is expected to form both magnetic and orbital order.

At filling one-eighth, the Fermi surface is within the bottom two spinful bands.
In the absence of $\pi$-bonding, the two bands become flat, which
enhances the interaction effect.
Due to the Coulomb interaction, the system favors a flat-band
ferromagnetic state~\cite{zhang2010proposed}.
Therefore, effectively, one of the spinful flat bands is filled,
and the Fermi surface is at the quadratic band touching point again.
The resulting weak interaction phase exhibits the anomalous quantum Hall effect~\cite{Chen_kagome2018}.

As the filling becomes even lower, the systems start to Wigner-crystallize.  When spin-orbit coupling is included, the flat band becomes nearly flat and acquires Chern number $\pm 1$.
In this case, the Chern fractional insulator~\cite{Sun2011} may become
a ground state candidate and competes with the Wigner crystal phase.


\section{Discussion and summary}
\label{sec:summary}
We have studied the orbital-active Dirac materials in a unified manner. The various orbital realizations can be understood as the irreps of the point group symmetry $C_{3v}$. All belonging to the two-dimensional $E$ irrep of $C_{3v}$, the $(p_x, p_y)$ doublet, the $(d_{xy}, d_{x^2-y^2})$ doublet and the $E_g$ doublet can be mapped to each other, and the Dirac materials based on these two sets of different doublet have the same universal properties.
Using $k\cdot p$ theory, we demonstrate that the symmetry leads to the
orbital enriched Dirac cone at $K(K^\prime)$ point and quadratic band
touching at the $\Gamma$ point.
The symmetry also enforces the unique orbital configuration of the wavefunction at these high symmetry points.
When only the $\sigma$-bonding is considered, the spectrum hosts
two flat bands, which can lead to exotic phases such as the
Wigner crystal.

Compared with other doublets, the $E_g$ doublet exhibits unique features. First, this doublet is naturally realized in a buckled honeycomb lattice instead of the planar one, leading to a distinct pattern of the Wigner crystal. Furthermore, in the $E_g$ doublet, the angular momentum is completely quenched, and the lowest order of the magnetic moment is the octupole moment. This leads to edge states carrying octupole moment once a topological gap is opened.

Orbital active Dirac materials are not only limited to the electronic systems but also include systems of phonons and polaritons~\cite{Jacqmin2014, Milicevic2017, Zhang2015, Roman2015, Stenull2016, Zhu2018}, where their polarization modes realize the orbital degrees of freedom.
The symmetry argument in Sec.~\ref{sec:general_symmetry} also enforces chiral valley phonons in materials with a honeycomb structure, such as boron nitride and transition metal dichalcogenides.
Thus the interplay between the chirality of electrons' wavefunction and the chirality of the phonons opens a new door for valleytronics.

Finally, we briefly discuss the band flatness of the Majorana fermions.
One dimensional Majorana edge modes can appear with flat dispersion
as protected by time-reversal symmetry~\cite{li2013spontaneous}.
The divergence of density of states leads to interesting interaction
effects by lifting band flatness via spontaneous time-reversal symmetry breaking.
Majorana modes with the cubic dispersion relation can also be realized
as the surface state with high topological index superconductivity
\cite{yang2016topological}.
Its density of states diverges at $k=0$, which can be viewed as
nearly flat.

\section{Declarations}

\textbf{Acknowledgments} -- C. W. is supported by the National Natural Science Foundation of China through Grant No. 12174317, No. 11729402 and No. 12234016.

\textbf{Contributions} -- C. W. initiated and supervised the project. S. X. and C. W. conducted research and wrote the manuscript. All authors read and approved the final manuscript.

\textbf{Competing interests
} -- The authors declare that they have no competing interests.

\textbf{Data, Material and/or Code availability} --
The data and code associated with the project are available upon reasonable request.

\textbf{Corresponding author} -- Correspondence to Congjun Wu.


%

\onecolumngrid

\appendix
\section{The $C_{3V}$ group and its double group $C_{3V}^D$}
\label{appendix:group}
The $C_{3v}$ point group is the simplest non-abelian group,
containing six elements generating by a three-fold rotation and an in-plane reflection. It has three irreps $A_1$, $A_2$ and $E$. The first two are one-dimensional while the last one is two-dimensional.  The $A_1$ irrep is trivial and examples include $s$ orbitals and $p_z$ orbitals; the $A_2$ irrep is odd under the reflection with realizations such as pseudovector $L_z$ and $f$ orbital $y(3x^2-y^2)$ . In this work, we are mostly interested in the two-dimensional $E$ irrep.

The $C_{3v}$ group includes 6 operations in 3 conjugacy classes:
the identity I, the 3-fold rotations
$\{C_3^1, C_3^2\}$ around the vertical axis, and the
reflection operations with respect to three vertical planes
$\{\sigma_{v_i}\}$ with $i=1\sim 3$.
It possesses two one-dimensional representations $A_1$ and $A_2$,
and one two-dimensional representation $E$.
Their character table is presented in Tab \ref{table_c3v}.
The bases of the $A_{1,2}$ representations carry angular momentum
quantum number $L_z=0$, and those of the $E$ representation
can be chosen with $L_z=\pm 1$.

\begin{table}[h]
\begin{center}
\begin{tabular}{|c|l|c|c|c|} \hline
 & I & 2$C_3$ & 3$\sigma_v$ \\ \hline
$A_1$& 1& 1& 1 \\ \hline
$A_2$& 1& 1& -1 \\ \hline
$E$ & 2& -1& 0 \\ \hline
\end{tabular}
\caption{The character table of the $C_{3v}$ group, which has
two one dimensional representations $A_{1,2}$ and one two-dimensional
representation $E$.
$A_{1,2}$ carry orbital angular momentum $L_z=0$, and
$E$ carries $L_z=\pm 1$.
}
\end{center}
\label{table_c3v}
\end{table}

In the presence of spin-orbit coupling, $C_{3v}$ is augmented to its
double group $C_{3v}^D=C_{3v} + \bar C_{3v}$.
$\bar C_{3v}=\bar I C_{3v}$ is the coset by multiplying $\bar I$
to $C_{3v}$, where $\bar I$ is the rotation of $2\pi$.
The $C^D_{3v}$ group has six conjugacy classes, and hence six non-equivalent
irreducible representations whose characteristic table is
presented in Tab. \ref{table_c3vd}.
$A_{1,2}$ and $E$ remain the representations of $C_{3v}^D$ of integer
angular momentum, for which $\bar I$ is the same as the identity
operation.
In addition, $C^D_{3V}$ also possesses half-integer angular momentum
representations, for which $\bar I$ is represented as the negative
of the identity matrix.
For example, a new two-dimensional representation $E_\frac{1}{2}$ appears
corresponding to the angular momentum $J_z=\pm\frac{1}{2}$.
The cases of $J_z=\pm\frac{3}{2}$ are often denoted as the
$E_{\frac{3}{2}}$ representation.
Actually, they are not an irreducible two-dimensional representation, but
two non-equivalent one-dimensional representations.
The two bases of $\psi_{J_z=\pm\frac{3}{2}}$ are equivalent under the 3-fold
rotations since $\frac{3}{2}\equiv -\frac{3}{2} (\mbox{mod}~ 3)$, and neither
of them are eigenstates of the reflections $\sigma_v$ and
$\bar\sigma_v=\bar I \sigma_v$.
Instead, their superpositions $\frac{1}{\sqrt 2} (\psi_\frac{3}{2}\pm
i\psi_{-\frac{3}{2}})$ carry the characters of $\pm i$ for $\sigma_v$
and $\mp i$ for $\bar \sigma_v$, respectively.

\begin{table}[h]
\begin{center}
\begin{tabular}{|c|c|c|c|c|c|c|} \hline
& I & $\bar I$ & $\{C^1_3, \bar C^2_3$ \} & $\{C^2_3, \bar C^1_3\}$ &
 3$\sigma_v$ & 3$\bar \sigma_v$ \\ \hline
$E_{\frac{1}{2}}$ & 2 &-2& 1 & -1 & 0 &0 \\ \hline
$E_{\frac{3}{2}}$ &1 &-1& -1 & 1 & $i$ & $-i$ \\
 &1 &-1& -1 & 1 &$-i$ & $i$ \\ \hline
\end{tabular}
\caption{Spinor representations for the $C_{3v}^D$ group:
The two-dimensional representation $E_{\frac{1}{2}}$ is of $J_z=\pm \frac{1}{2}$.
$E_{\frac{3}{2}}$ splits into two non-equivalent one-dimensional
representations with different characters under vertical
reflections.
}
\end{center}
\label{table_c3vd}
\end{table}

\section{Spherical tensor operators in the $d$-orbital space}
\label{appendix:tensor}
In the Hilbert space of $d$ orbitals, the angular momentum operators are defined in the standard way
\bea
\hat  L_x  =
\left(
\begin{array}{ccccc}
 0 & 1 & 0 & 0 & 0 \\
 1 & 0 & \sqrt{\frac{3}{2}} & 0 & 0 \\
 0 & \sqrt{\frac{3}{2}} & 0 & \sqrt{\frac{3}{2}} & 0
   \\
 0 & 0 & \sqrt{\frac{3}{2}} & 0 & 1 \\
 0 & 0 & 0 & 1 & 0 \\
\end{array}
\right) ,  \ \ \
\hat L_y=
\left(
\begin{array}{ccccc}
 0 & -i & 0 & 0 & 0 \\
 i & 0 & -i \sqrt{\frac{3}{2}} & 0 & 0 \\
 0 & i \sqrt{\frac{3}{2}} & 0 & -i
   \sqrt{\frac{3}{2}} & 0 \\
 0 & 0 & i \sqrt{\frac{3}{2}} & 0 & -i \\
 0 & 0 & 0 & i & 0 \\
\end{array}
\right) , \ \ \
\hat L_z=
\left(
\begin{array}{ccccc}
 2 & 0 & 0 & 0 & 0 \\
 0 & 1 & 0 & 0 & 0 \\
 0 & 0 & 0 & 0 & 0 \\
 0 & 0 & 0 & -1 & 0 \\
 0 & 0 & 0 & 0 & -2 \\
\end{array}
\right) \\
\eea
The total angular momentum operator $\hat L^2 =\hat L_x^2+\hat L_y^2+\hat L_z^2$, and the ladder operators $\hat L_\pm=\hat L_x \pm  \hat L_y$. The spherical tensors $\hat Y_{lm}$ satisfy the following commutation relation,
\bea
\ [ L_+, Y_{l,m} \ ]=\sqrt{(l-m)(l+m+1)}Y_{l,m+1}
\eea
Fixing $l$, the tensor operator with the lowest $m$ can be easily expressed as powers of $\hat L_-$,
\bea
\hat Y_{l,-l}=\frac{\sqrt{(2l) !}}{2^l l!}(\hat L_-)^l
\eea
Based on these relations, the general rank $l$ spherical tensors can be constructed systematically from the angular momentum operators.
All 25 linear independent operators acting on $d$ orbitals can be organized into spherical tensor operators with rank $0\sim 4$.  The rank 1 tensor operators are,
\bea
\hat Y_{1,-1}=\frac{1}{\sqrt{2}}\hat L_- , \ \ \
\hat Y_{1,0}=\hat L_z, \ \ \
\hat Y_{1,1}=-\frac{1}{\sqrt{2}} \hat L_+
\eea
The rank 2 tensor operators are
\bea
\hat Y_{2,-2}=\sqrt{\frac{3}{8}} \hat L_-^2 , \quad \hat Y_{2,-1}=\sqrt{\frac{3}{2}} \overline{\hat L_- \hat L_z}, \quad
\hat Y_{2,0}=\frac{1}{2} (2\hat L_z^2-\hat L_x^2-\hat L_y^2), \quad
\hat Y_{2,1}=-\sqrt{\frac{3}{2}} \overline{\hat L_+ \hat L_z}, \ \ \
\hat Y_{2,2}=\sqrt{\frac{3}{8}} \hat L_+^2
\eea
where bars over the operators represent the average over all possible  operators ordering. The rank three tensor operators are
\bea
&\hat Y_{3,-3}=\frac{\sqrt 5}{4} \hat L_-^3, \quad \hat Y_{3,-2}=\sqrt{\frac{15}{8}}  \overline{\hat L_-^2 \hat L_z}, \quad \hat Y_{3,-1}=\frac{\sqrt{3}}{4}\overline{\hat L_-(4L_z^2-L_x^2-L_y^2)}, \quad
\hat Y_{3,0}=\frac{1}{2} \overline{L_z (2\hat L_z^2-3 L_x^2-3L_y^2)} \\
&\hat Y_{3,1}=-\frac{\sqrt{3}}{4} \overline{\hat L_+(4L_z^2-L_x^2-L_y^2)},
\quad\hat Y_{3,3}=\sqrt{\frac{15}{8}}  \overline{\hat L_+^2 \hat L_z}, \quad \hat Y_{3,-3}=-\frac{\sqrt 5}{4} \hat L_+^3.
\eea
Lastly, the rank 4 tensors are
\bea
&\hat Y_{4,-4}=\sqrt{\frac{35}{128}} \hat L_-^4, \quad
\hat Y_{4,-3}=\frac{\sqrt {35}}{4} \overline{\hat L_-^3 \hat L_z}, \quad
\hat Y_{4,-2}=\sqrt{\frac{5}{32}} \overline{\hat L_-^2 (7 \hat L_z^2-\hat L^2)} \quad \hat Y_{4,-1}=\frac{\sqrt 5}{4}\overline{\hat L_-\hat L_z (7\hat L_z^2-3\hat L^2)} \\
&\hat Y_{4,0}=\frac{1}{8} \overline {(35 \hat L_z^4-30 \hat L_z^2 \hat L^2+3 \hat L^4)},\quad \hat Y_{4,1}=-\frac{\sqrt 5}{4}\overline{\hat L_+\hat L_z (7\hat L_z^2-3\hat L^2)},\quad \hat Y_{4,2}=\sqrt{\frac{5}{32}} \overline{\hat L_+^2 (7 \hat L_z^2-\hat L^2)} \\
&\hat Y_{4,3}=-\frac{\sqrt {35}}{4} \overline{\hat L_+^3 \hat L_z}, \quad  \hat Y_{4,4}=\sqrt{\frac{35}{128}} \hat L_+^4.
\eea

\end{document}